%% file: TSP_V3.tex
\newif\ifsingle

\newif\ifFullVersion

\ifsingle
\documentclass[10pt,onecolumn,draftclsnofoot]{IEEEtran}
\else		
\documentclass[10pt,final, twocolumn]{IEEEtran}
\fi

\usepackage{times}
\usepackage{amsmath,dsfont}
\usepackage{amssymb}
\usepackage{epsfig} 
\usepackage{setspace}
\usepackage{color}
\usepackage{cite}
\usepackage{epstopdf}
\usepackage{subfigure}
\usepackage{graphics}
\usepackage{accents}
\usepackage{acronym}
\usepackage[bookmarks,colorlinks]{hyperref}
\usepackage{mathtools}
\usepackage{algorithm}
\usepackage{caption}
\usepackage{algcompatible}
\usepackage{enumitem}
\usepackage[most]{tcolorbox}
\usepackage{cuted}

\newtheorem{prop}{Proposition}

\newcommand{\myVec}[1]{{\boldsymbol{#1}}}

\definecolor{mypurple}{rgb}{0.910, 0.910, 0.969}
\definecolor{myblue}{rgb}{0.122, 0.435, 0.698}

\acrodef{fl}[FL]{federated learning}
\acrodef{dnn}[DNN]{deep neural network}
\acrodef{iot}[IOT]{internet-of-things}
\acrodef{sgd}[SGD]{stochastic gradient descent}
\acrodef{mimo}[MIMO]{multiple-input multiple-output}

\ifFullVersion
\newcommand{\myCite}[1]{~\cite{#1}}
\else
\newcommand{\myCite}[1]{}
\fi

\input{macros}

\IEEEoverridecommandlockouts 

\renewcommand{\baselinestretch}{0.97}

\begin{document}

\title{Federated Learning from Heterogeneous Data via Controlled Bayesian Air Aggregation
}

\author{
  {Tomer Gafni, Kobi Cohen, Yonina C. Eldar}
	\thanks{
		T. Gafni and K. Cohen are with the School of Electrical and Computer Engineering, Ben-Gurion University of the Negev, Beer-Sheva, Israel (e-mail:gafnito@post.bgu.ac.il; yakovsec@bgu.ac.il). Yonina C. Eldar is with the Math and CS Faculty, Weizmann Institute of Science, Rehovot, Israel (e-mail: yonina.eldar@weizmann.ac.il).
	}
	\thanks{This work was supported by the Israel Science Foundation under Grant 2640/20. A short version of this paper that introduces the COBAAF algorithm, and preliminary simulation results was presented in Proceedings of the 58th Annual Allerton Conference on Communication, Control, and Computing (Allerton) \cite{gafni2022cobaaf}. In this journal version we include: (i) A detailed discussion on the implementation of the BAAF and COBAAF algorithms; (ii) a deep theoretical analysis of the algorithms with detailed proofs; (iii) more extensive simulation results; and (iv) a detailed discussion of the results, and comprehensive discussion and comparison with the existing literature.}
	\vspace{-0.75cm}
}
\maketitle

\begin{abstract}
\label{sec:abstract}
Federated learning (FL) is an emerging machine learning paradigm for training models across multiple edge devices holding local data sets, without explicitly exchanging the data. Recently, over-the-air (OTA) FL has been suggested to reduce the bandwidth and energy consumption, by allowing the users to transmit their data simultaneously over a Multiple Access Channel (MAC). 
However, this approach results in channel noise directly affecting the optimization procedure, which may degrade the accuracy of the trained model. 
In this paper we jointly exploit the prior distribution of local weights and the channel distribution, and develop an OTA FL algorithm based on a Bayesian approach for signal aggregation. Our proposed algorithm, dubbed Bayesian Air Aggregation Federated learning (BAAF), is shown to effectively mitigate noise and fading effects induced by the channel.
To handle statistical heterogeneity of users data, which is a second major challenge in FL, we extend BAAF to allow for appropriate local updates by the users and develop the Controlled Bayesian Air Aggregation Federated-learning (COBAAF) algorithm.
In addition to using a Bayesian approach to average the channel output, COBAAF controls the drift in local updates using a judicious design of correction terms. 
We analyze the convergence of the learned global model using BAAF and COBAAF in noisy and heterogeneous environment, showing their ability to achieve a convergence rate similar to that achieved over error-free channels. Simulation results demonstrate the improved convergence of BAAF and COBAAF over existing algorithms in machine learning tasks.
\end{abstract}

\section{Introduction}
\label{sec:Introduction}
In recent years, machine learning algorithms have given rise to advances in multiple fields, such as natural language processing, computer vision, and speech recognition \cite{lecun2015deep}. Since these algorithms are data driven, their success hinges on vast amounts of training data. Nowadays, a lot of this data is generated on edge devices such as mobile phones, sensors, vehicles, and medical devices. The common strategy is to gather these samples at a computationally powerful server, which uses them to train its model \cite{chen2019deep}. Due to privacy and locality concerns, uploading of data sets, such as images and text messages, may be undesirable. Furthermore, the communication link between edge devices and the server can be heavily burdened by sharing of massive data sets.

These difficulties can be tackled by exploiting the increased computational power of mobile devices via mobile edge computing, in which training and inference is performed directly at network edges without having to share the raw data. In such systems, learning from distributed data and communicating between the edge server and devices are two critical and coupled aspects. This motivates the edge learning framework of federated learning (FL), which features distributed learning at edge devices with centralized aggregations, orchestrated by an edge server, and is the focus of growing research attention over the last few years \cite{mcmahan2017communication, gafni2022federated, kairouz2021advances}. 
The fact that the complete data set is not available to the server leads to several core challenges that are not encountered in conventional machine learning methods. In this paper we focus on two of the main challenges in FL: Noise induced by the shared channel due to the communication bottleneck, and statistical heterogeneity of users' data. 

The communication bottleneck in FL arises from the need for a large number of users to repeatedly and concurrently send their intermediate learned models (e.g., deep neural networks weights) to the server over shared and resource-limited wireless networks. 
A commonly adopted strategy to tackle this challenge is to model the wireless channel as a set of independent error-free bit-limited links between the users and the server, e.g., by using frequency division multiplexing (FDM), and then to apply methods of compressing the conveyed model parameters via quantization \cite{alistarh2017qsgd, reisizadeh2020fedpaq,shlezinger2020federated} or sparsification \cite{aji2017sparse, alistarh2018convergence}. 
This, however, results in each user being assigned a dedicated bandwidth which decreases with the number of participating users, increasing the energy consumption required to meet a desirable communication rate, and decreasing in the overall throughput and training speed.
An alternative approach is to allow the users to simultaneously utilize the complete temporal and spectral resources of the uplink Multiple Access Channel (MAC) in a non-orthogonal manner, and in this way exploit the inherent aggregation carried out by the shared channel as a form of over-the-air (OTA) computation \cite{liu2020over, mergen2006type, Mergen_Asymptotic_2007, Liu_Type_2007, Marano_Likelihood_2007, cohen2013performance, cohen2019time}. 

Recently, OTA has been studied in the FL context \cite{sery2020analog, amiri2020machine, seif2020wireless, guo2020analog, frey2021over, zhang2021gradient, narayanamurthy2022federated, cao2022distributed, mu2022optimizing}.
This approach allows each transmitter to utilize the complete available channel resources, e.g., bandwidth, regardless of the number of users.
However, this results in \emph{channel noise} directly affecting the model weights. This may degrade the accuracy of the trained model as SGD is known to be sensitive to noisy observations \cite{cesa2011online}.
In \cite{amiri2020machine}, the authors considered scenario where the model updates are sparse with an identical sparsity pattern, which is not likely to hold when the data is heterogeneous.
In \cite{leebayesian,park2021bayesian}, the authors used a Bayesian aggregation procedure to handle the channel noise and fading. However, they considered distributed application of full gradient descent optimization over noisy channels with a single local step at each round, which is less efficient in terms of communication and computation resources as compared to FedAvg. 

In our previous work, we proposed the COTAF algorithm \cite{sery2021over} which facilitates high throughput FL over wireless channels, while preserving the convergence properties of the common FedAvg method for distributed learning, by introducing a time-varying precoding at the users. 
While COTAF has demonstrated some success in handling noise, in this work we take an additional step towards better noise reduction. By exploiting the prior distribution of local weights and the channel distribution, we address the noise challenge in OTA FL using a Bayesian approach, introducing the Bayesian Air Aggregation Federated-learning (BAAF) algorithm. Similarly to previous OTA FL methods, BAAF utilizes the fact that the model updates in FL are typically transmitted to be averaged, and uses OTA transmissions for increasing the throughput when operating over a shared channel. The transmission is then judiciously designed such that the received noisy signal at the server follows a Bayesian combining strategy. Specifically, we jointly exploit the distribution of the local models and the communication channels when aggregating the noisy local models received by the server, and estimate the noise-free sum of local models. Under the premise that the prior distributions of the local models are jointly Gaussian and the channel is Gaussian with block fading, this aggregation method is optimal in the sense of minimizing the mean-squared error (MSE) per communication round.

We analyze theoretically the convergence of models trained by BAAF, when the objective is a strongly convex and smooth function with bounded gradients, as commonly assumed for purposes of theoretical analysis in FL convergence studies \cite{li2019convergence,stich2018local}. We rigorously show the effect of our MMSE estimation on the convergence of the algorithm compared to FedAvg that assumed noise-free and fading-free channels, and show superior theoretical performance over \cite{sery2021over} in handling the noise under the same statistical model.
Specifically, BAAF achieves the same convergence rate as FedAvg, with an additional noise term which depends on the MSE of our aggregation strategy and decays as $1/R$, where $R$ is the number of communication rounds. We show explicitly that our error term convergences to zero with a higher rate compared to the corresponding term in \cite{sery2021over}.
We also analyze the performance of BAAF numerically, and demonstrate its improved performance over conventional OTA FL and over the COTAF algorithm.

Our second contribution is the extension of our results to FL with statistical heterogeneity among users' data sets.
Statistical heterogeneity implies that the data generating distributions vary among different sets of users \cite{li2019convergence}. This is often the case in FL, as the data available at each user's device is likely to be personalized. Statistical heterogeneity implies that when training several instances of a model on multiple edge devices, each instance may be biased.
While FedAvg \cite{mcmahan2017communication} is a very popular algorithm for FL, implementing it on heterogeneous data results in slow and unstable convergence due to the drift in the local updates.
Several solutions have been suggested to address statistical heterogeneity, such as adding a regularizer to ensure users updates remain close \cite{li2020federated}, using a primal-dual optimization perspective \cite{zhang2021fedpd}, and more \cite{wang2021novel, khaled2020tighter, khaled2019first, cen2020convergence, fallah2020personalized, karimireddy2020mime, karimireddy2020scaffold, li2019convergence, ramazanli2020adaptive, woodworth2020minibatch, acar2021federated}.

In this work we adopt the technique used in the SCAFFOLD algorithm \cite{karimireddy2020scaffold}, which uses \emph{control variates} to reduce the user-drift and ensure that the user updates are aligned with each other. Specifically, SCAFFOLD estimates the update direction (i.e., gradient) for the server model ($c$)  and the update direction for each user $c_i$ referred to as a control variant. The difference ($c-c_i$) is then an estimate of the user-drift which is used to correct the local update, as we elaborate on Section \ref{sec:CoBAAF}. This strategy successfully overcomes heterogeneity and converges in a significantly smaller number of communication rounds compared to FedAvg. However, SCAFFOLD is implemented in an ideal scenario assuming that the received local models at the server are error-free. 
Here, we implement the technique of control variates over noisy channels, and analyze the trade-off between the errors corresponding to heterogeneity and channel noise on the convergence rate. To handle statistical heterogeneity of users data, we extend BAAF and develop the COBAAF algorithm. 
First, we use control variates in the local updates of the users to address the drift induced by the statistical heterogeneity of the data. Then, similarly to BAAF, we use OTA transmission for both models and controls variates with an appropriate precoding, and implement minimum mean-square error (MMSE) estimation to mitigate the noise effect. Finally, the estimated global model and control vairates are distributed to the users for another round.

We analyze the convergence of models trained by COBAAF in heterogeneous and noisy environments to the minimal achievable loss function. We rigorously show the effect of our MMSE aggregation on the convergence of the algorithm compared to SCAFFOLD \cite{karimireddy2020scaffold} that assumed noise-free and fading-free channels, and show superior performances over COTAF \cite{sery2021over} in terms of coping with statistical heterogeneity. 
We further exploit an important trade-off between the channel error (determined by the MSE of the Bayesian estimation) and the heterogeneity error (determined by the gradient dissimilarity parameter).
Finally, we analyze the performance of COBAAF numerically, and demonstrate its improved convergence compared to existing algorithms in simulation results.

The rest of this paper is organized as follows: Section \ref{sec:problem} briefly reviews the FedAvg algorithm and some of its relevant extensions, and presents the system model of Bayesian OTA FL. Section \ref{sec:Bayesian_Aggregation} presents the BAAF algorithm along with its theoretical convergence analysis. In Section \ref{sec:CoBAAF} we present the extension to the heterogeneous setting and present the COBAAF algorithm along with its convergence analysis. Numerical results are described in Section \ref{sec:Simulations}. Finally, Section \ref{sec:conclusions} provides concluding remarks. Detailed proofs of our main results are given in the appendix.

Throughout the paper, we use boldface lower-case letters for vectors, e.g., $\myVec{x}$. The $\ell_2$ norm, stochastic expectation, and Gaussian distribution are denoted by $||\cdot||$, $\mathbb{E}[\cdot]$, and $\mathcal{N}(\cdot,\cdot)$, respectively. Finally, $\myVec{I}_n$ is the $n \times n$ identity matrix, and $\mathbb{R}$ is the set of real numbers.
\color{black}

\section{System Model and Problem Formulation}
\label{sec:problem}
In this section we detail the system model for which BAAF is derived.  In Section \ref{sec:CoBAAF} we extend BAAF to cope with the heterogeneous nature of FL, and present the COBAAF algorithm. 

\subsection{Federated Learning Architecture}
\label{ssec:Federated Learning}
We consider a set of $N$ users, indexed by $\mathcal{N}\triangleq\{1,\ldots,N\}$. Each user (say user $i \in \mathcal{N}$) has access to a local data set $\mathcal{D}_i$ of $D_i$ entities, denoted by $\{\myVec{s}_i^n\}_{n=1}^{D_i}$, sampled in an i.i.d. fashion from $\mathcal{D}_i$, where $s_i^n$ consists of the training input and the corresponding label. We assume that the data between users is heterogeneous, i.e., the local distributions $\{\mathcal{D}_i \}$ are non-identical.
Each user uses its data set to train a local model comprised of $d$ parameters, represented by the vector $\myVec{\theta} \in \Theta \subset \mathbb{R}^d$. Training is carried out  to minimize a local objective $f_i(\myVec{\theta})$, based on a loss measure $l(\cdot ; \myVec{\theta})$. The local objective for user $i$ is given by:
\begin{equation}
\label{eq:local_loss}
    \displaystyle f_i(\myVec{\theta}) = \frac{1}{D_i}\sum \limits_{n=1}^{D_i} l(\myVec{s}_i^n ; \myVec{\theta} ).     
\end{equation}
Consequently, the objective of user $i$ is to recover the parameters $\boldsymbol{\theta}_i$ that minimize \eqref{eq:local_loss}:
\begin{equation}
\label{eq:local_objective}
\displaystyle \boldsymbol{\theta}_{i}^* = \arg\min_{{\boldsymbol{\theta} }} f_i(\myVec{\theta}). 
\end{equation}

 The objective of the server is to utilize the data available at the users side to train a global model with parameters $\boldsymbol{\theta}$. The objective used for learning $\boldsymbol{\theta}$ is given by:
\begin{equation}
\label{eq:global_loss}
    F(\boldsymbol{\theta}) = \frac{1}{N} \sum \limits_{i=1}^N f_i(\myVec{\theta}).
\end{equation}
Consequently, the server aims to solve the following minimization problem:
\begin{equation}
\label{eq:global_objective}
    \displaystyle \boldsymbol{\theta}^* = \arg\min_{\boldsymbol{\theta}} F(\boldsymbol{\theta}).
\end{equation}

\subsection{Federated Averaging}
\label{ssec:FedAvg}
Federated Averaging (FedAvg) is a distributed learning algorithm aimed at recovering (\ref{eq:global_objective}), without having the users share their local data.
We first present the algorithm when executed over an error-free channel.
The iterative algorithm carries out multiple training rounds, each consisting of the following three phases:
\begin{enumerate} 
    \item In global round $r$, starting from the shared model $\myVec{\theta}^r_{i,0}=\myVec{\theta}^{r-1}$, each user $i$ trains its local model using its local data set over $K$ SGD local steps, namely, 
    \begin{equation}
        \label{eq:local_SGD_Iteration}
        \myVec{\theta}_{i,k}^r = \myVec{\theta}_{i,k-1}^r - \eta \myVec{g}_i(\myVec{\theta}_{i,k-1}^r), k=1,\ldots,K,
    \end{equation}
    where $\myVec{g}_i(\myVec{\theta}_{i,k-1}^r) = \nabla f_i(\myVec{\theta}_{i,k-1}^r,\xi_{i,k-1})$ is an unbiased stochastic gradient of $f_i$ at local step $k$ satisfying $\mathbb{E}_{\xi}[\myVec{g}_i(\myVec{\theta})] = \nabla f_i(\myVec{\theta})$, and $\eta$ is the step size. 
    \item Each user $i$ conveys its trained local model $\myVec{\theta}_{i,K}^r$ (or alternatively, the updates in its trained model $\myVec{\theta}_{i,K}^r - \myVec{\theta}_{i,0}^{r}$) to the central server, which averages them into a global model via $\myVec{\theta}^r = \frac{1}{N} \sum_{i=1}^N \myVec{\theta}_{i,K}^r
    $.
    \item The server shares its new model with the users for the next round.
\end{enumerate}

The uplink transmission in this algorithm is typically executed over an error-free channel with limited throughput, where channel noise and fading are assumed to be eliminated. In this paper, we consider transmissions in practical wireless networks, and we thus address the aggregated scheme in the present of channel noise and channel fading, as detailed in the sequel.

\subsection{Statistical and Communication Channel Model}
\label{ssec:Channel}

As commonly assumed in OTA FL studies (see e.g.,  \cite{sery2021over,amiri2020machine,cao2022distributed, paul2021accelerated}), the $N$ users transmit their data to the server via a MAC, since the uplink communication is typically notably more constrained as compared to its downlink counterpart in terms of throughput. The downlink channel used for server transmissions is modeled as supporting reliable communications at arbitrary rates. 

We next formulate the uplink channel model. Each user conveys its local model $\myVec{\theta}_i^r \triangleq \myVec{\theta}_{i,K}^r$ to a channel input $\myVec{x}_i^r$ (which we define precisely in Section \ref{ssec:Full}).
Wireless channels are inherently a shared and noisy media.
Hence, each user of index $i$ experiences at round $r$ a block fading channel $\tilde{h}_i^r = h_i^r e^{j \phi_i^r}$, where $h_i^r>0$ and $e^{j \phi_i^r} \in [-\pi, \pi]$ are its magnitude and phase, respectively. The channel output received by the server at communication round $r$ when each user transmits a $d \times 1$ vector $\myVec{x}_i^r$, is thus given by
\begin{equation}
\label{eq:channel_model}
  \Tilde{\myVec{y}}^r = \sum \limits_{i=1}^N \tilde{h}_i^r \myVec{x}_i^r +\myVec{\Tilde{w}}^r,
\end{equation}
where $\myVec{\Tilde{w}}^r \sim \mathcal{N}(0,\sigma_w^2 \myVec{I}_d)$ is a $d \times 1$ vector of additive noise. 
The channel input is subject to an average power constraint: 
\begin{equation}
\label{eq:energy_constraint}
  \mathbb{E}[||\myVec{x}_i^r||^2] \leq P,
\end{equation}
where $P>0$ represents the available transmission power.

To simplify the presentation, we start by assuming a unit fading, i.e., $\tilde{h}_i^r=1$ for all $r, i$. In Section \ref{ssec:fading_channels} we extend the design and analysis to transmissions over fading MAC with general fading values.

We now present the statistical framework for our aggregation scheme. The motivation of our proposed algorithm is to find an optimal aggregation method of local weights from a Bayesian viewpoint.
Characterizing the prior distribution of local weights is very challenging because it not only depends on the loss function but also on the underlying distribution of data samples. To overcome this difficulty, we model the distribution of the local weights as a Gaussian prior distribution with a proper moment matching technique.
Specifically, we model the $d$ elements of $\myVec{\theta}_i^r = [\theta_i^{r,1},\ldots \theta_i^{r,d}]$ as i.i.d Gaussian random variables with common mean $\mu_i^r$ and variance $(\sigma_i^r)^2$, which can be computed by performing offline simulations over a smaller data set, and therefore can be stored prior to the execution of the algorithm.
The prior distribution can change over communication rounds according to the underlying distribution of data samples and the local loss function.
We assume that $\myVec{\theta}_i^r$ and $\myVec{\theta}_n^r$ are uncorrelated for $i \neq n$, as assumed in \cite{leebayesian}.
\color{black}

\section{The Bayesian Air Aggregation Federated-learning (BAAF) Algorithm}
\label{sec:Bayesian_Aggregation}

\subsection{Algorithm Design}
\label{ssec:Full}
We now present the proposed BAAF algorithm.
First, we adapt the method proposed in \cite{sery2021over}, where each user first precodes its model update $\myVec{\theta}_i^r - \myVec{\theta}_{i,0}^{r}$
into the MAC channel input $\myVec{x}_i^r$ via:
\begin{equation}
\label{eq:channel_input}
    \myVec{x}_i^r = \sqrt{\alpha^r}(\myVec{\theta}_i^r - \myVec{\theta}_{i,0}^{r}),
\end{equation}
where $\sqrt{\alpha^r}$ is a precoding factor set to gradually amplify the model updates as $r$ progresses, while satisfying the power constraint (\ref{eq:energy_constraint}). The precoder $\sqrt{\alpha^r}$ is given by
\begin{equation}
\label{eq:precoder}
    \alpha^r \triangleq \frac{P}{\max_i \mathbb{E}[||\myVec{\theta}_i^r - \myVec{\theta}_{i,0}^{r}||^2]}.
\end{equation}
The precoding parameter $\sqrt{\alpha^r}$ depends on the distribution of the updated model, which depends on the distribution of the data. It can thus be computed by performing offline simulations with smaller data sets and distributing the numerically computed coefficients among the users, as we do in our numerical study in Section \ref{sec:Simulations}. Alternatively, when the loss function has bounded gradients, this term can be replaced with a coefficient that is determined by the bound on the norm of the gradients (see \cite{sery2021over}). 

In each communication round, the server thus receives the \textit{noisy} sum of local model updates:
\begin{equation}
\label{eq:updates_channel_output}
\Tilde{\myVec{y}}^r = \sum \limits_{i=1}^N \sqrt{\alpha^r}(\myVec{\theta}_i^r - \myVec{\theta}_{i,0}^{r}) +\myVec{\Tilde{w}}^r.
\end{equation}
In order to recover the aggregated global model $\myVec{\theta}^r$ from $\Tilde{\myVec{y}}^r$, the server sets:
\begin{equation}
\label{eq:output_processing}
    \Tilde{\myVec{\theta}}^r = \frac{\Tilde{\myVec{y}}^r}{N \sqrt{\alpha^r}} + \myVec{\theta}^{r-1},
\end{equation}
which can be equivalently written as:
\begin{equation}
\label{eq:output_processing_eq}
    \Tilde{\myVec{\theta}}^r = \frac{1}{N} \sum \limits_{i=1}^N \myVec{\theta}_i^r +\myVec{w}^r,
\end{equation}
where $\myVec{w}^r \triangleq \frac{\Tilde{\myVec{w}}^r}{N \sqrt{\alpha^r}}$ is the equivalent additive noise term distributed via $\myVec{w}^r \sim \mathcal{N}(0,\frac{\sigma_w^2}{N^2 \alpha^r} \myVec{I}_d)$.

Finally, the server aims to estimate the average $\myVec{\theta}^r = \frac{1}{N}\sum_{i=1}^N \myVec{\theta}_i^r$. Our goal is thus to find a function $U(\Tilde{\myVec{\theta}}^r)$ such that 
\begin{equation}
\label{eq:MSE_Criterion}
    \hat{\myVec{\theta}}_{\text{MMSE}}^r = \arg \min_U \mathbb{E}\left[||\myVec{\theta}^r - U(\Tilde{\myVec{\theta}}^r)||_2^2\right].
\end{equation}
Let $\myVec{\hat{\theta}}^r_{\text{MMSE}} \triangleq [\hat{\theta}^{r,1}_{\text{MMSE}}, \ldots, \hat{\theta}^{r,d}_{\text{MMSE}}]^T$, and $\myVec{\Tilde{\theta}}^r \triangleq [\Tilde{\theta}^{r,1}, \ldots, \Tilde{\theta}^{r,d}]^T$.
The MSE-optimal function 
for the optimization problems in (\ref{eq:MSE_Criterion}) is obtained by the conditional expectation \cite{stark1986probability}:
\begin{align}
\label{eq:MSE_Solution}
    \hat{\theta}_{\text{MMSE}}^{r,m} = \mathbb{E}\left[\theta^{r,m} | \Tilde{\myVec{\theta}}^r \right] = \mu^r  + \frac{(\sigma^r)^2}{(\sigma^r)^2+ \frac{\sigma_w^2}{\alpha^r N^2}} (\Tilde{\theta}^{r,m} - \mu^r),
\end{align}
where $ \mu^{r} \triangleq \frac{1}{N} \sum \limits_{i=1}^N \mu_i^{r}$ and $(\sigma^{r})^2 \triangleq \frac{1}{N^2} \sum \limits_{i=1}^N (\sigma_i^{r})^2$. Eq.~(\ref{eq:MSE_Solution}) is proved in Appendix \ref{sec:App_Lemma_1_2}.
We note that in order to evaluate the estimator in (\ref{eq:MSE_Solution}), the server requires to know $\mu_i^r$, $(\sigma_i^r)^2$ and $\sigma_w^2$. User $i$ can send these parameters using a powerful channel code (e.g., polar codes) so that the server perfectly decodes them. When $\myVec{\theta}_i^r$ and $\myVec{\theta}_n^r$ are correlated or when the elements of $\myVec{\theta}_i^r$ are not statistically identical, the estimation in (\ref{eq:MSE_Solution}) is suboptimal. Nevertheless, BAAF is still a promising algorithm due to its computational scalability in both the number of mobile devices and the model size at the cost of an MSE performance loss, and numerical experiments presented in Section \ref{sec:Simulations} indicate that BAAF indeed mitigates effectively the channel noise.

After computing the estimated sum in (\ref{eq:MSE_Solution}), the server  sends back $\myVec{\hat{\theta}}_{\text{MMSE}}^r$ to the users, and the local model is updated by: $\myVec{\theta}^{r+1}_{i,0} = \myVec{\hat{\theta}}_{\text{MMSE}}^r$. 
The local updated rule of user $i$ at round $r+1$ under BAAF is then given by:
\begin{equation}
    \label{real_local_update}
\myVec{\theta}_{i,k}^{r+1} = \myVec{\theta}_{i,k-1}^{r+1} - \eta_l \myVec{g}_i(\myVec{\theta}_{i,k-1}^{r+1}), k=1,\ldots,K.    
\end{equation}
The final output of the algorithm is a weighted average for some positive weights (defined explicitly in Appendix \ref{sec:App_Th_1}) $\{w_r\}$ for $r \in \{1,\ldots,R+1\}$
\begin{equation}
\label{eq:final_output}
    \myVec{\bar{ \theta}}^R = \frac{1}{\sum_r w_r} \sum_r w_r \myVec{\hat{\theta}}_{\text{MMSE}}^{r-1}.
\end{equation}
The pseudocode of BAAF is given in Algorithm \ref{Algorithm1}.

\begin{algorithm}
\caption{BAAF Algorithm} 
\label{Algorithm1}
  \begin{algorithmic}[1]
    \STATE  \textbf{server input} initial parameter $\myVec{\theta}$
    \STATE  \textbf{user input} local step size $\eta_l$, means $\mu_i^r$, variances $(\sigma_i^r)^2$
    \FOR{each communication round $r=1,\ldots,R$} 
        \STATE     communicate ($\myVec{\hat{\theta}}^{r-1}_{\text{MMSE}}$) to all users
        \FOR{each user $i=1,\ldots,N$} 
            \STATE   initialize local parameters $\myVec{\theta_{i}^r} \leftarrow \myVec{\hat{\theta}}^{r-1}_{\text{MMSE}}$
            \FOR{each local step $k=1,\ldots,K$}
                \STATE compute a stochastic gradient $g_i(\myVec{\theta}_i^r)$ of $f_i$
                \STATE $\myVec{\theta_i^r} \leftarrow \myVec{\theta_i^r}-\eta_l g_i(\myVec{\theta}_i^r)$
            \ENDFOR
            \STATE  communicate $\left(\myVec{\theta}_i^r,\mu_i^r,(\sigma_i^r)^2\right)$
        \ENDFOR
        \STATE recover $\Tilde{\myVec{\theta}}^r$ according to (\ref{eq:output_processing})
        \STATE estimate $\hat{\theta}^{r,m}_{\text{MMSE}} = \mu^r  + \frac{(\sigma^r)^2}{(\sigma^r)^2+ \frac{\sigma_w^2}{\alpha^r N^2}} (\Tilde{\theta}^{r,m} - \mu^r)$        
    \ENDFOR
  \end{algorithmic}
\end{algorithm}

\subsection{Theoretical Analysis}
\label{ssec:Performance_Analsis}
We next analyze theoretically the convergence of BAAF to the optimal model parameters $\myVec{\theta}^*$, i.e., the vector $\myVec{\theta}$ which minimizes the global loss function. Our analysis is carried out under standard strong-convexity, smoothness, and bounded gradients assumptions used in SGD and FL studies (see e.g., \cite{li2019convergence,stich2018local,sery2021over,amiri2020machine}): \\
\textit{AS1}: Each function $f_i$ is $\beta$-smooth and $\mu$-strongly convex. \\
\textit{AS2}: The unbiased stochastic gradients $g_i(\myVec{\theta})$ satisfy $\mathbb{E}_{\xi}[||g_i(\myVec{\theta})-\nabla f_i(\myVec{\theta}) ||^2] \leq \sigma^2$ and $\mathbb{E}_{\xi}[||g_i(\myVec{\theta})||^2] \leq M^2$ for some fixed $\sigma^2>0$ and $M^2>0$, for each $\myVec{\theta} \in \Theta$ and $i \in \mathcal{N}$.\\ 
\textit{AS3}: There exist constants $G \geq 0$ and $B \geq 1$ such that:
\begin{equation*}
    \frac{1}{N} \sum \limits_{i=1}^N ||\nabla f_i(\myVec{\theta})||^2 \leq G^2 + B^2 ||\nabla f(\myVec{\theta})||^2, \; \forall \myVec{\theta} \in \Theta.
\end{equation*}

We next present the resulted MMSE estimation error.

\begin{prop}
    \label{lemma:MSE}
    The minimum MSE, $\text{MSE}(\hat{\theta}^{r,m}_{\text{MMSE}}) \triangleq \mathbb{E} \left[(\theta^{r,m} - \hat{\theta}^{r,m}_{\text{MMSE}})^2 \right]$, is given by:
    \begin{equation}
    \label{eq:LemmaMSE}
    \text{MSE}(\hat{\theta}^{r,m}_{\text{MMSE}}) = \frac{1}{N^2 \alpha^r} \cdot \frac{\sigma_w^2 }{1 + \frac{\sigma_w^2}{\alpha^r N^2 (\sigma^{r})^2}}  = \frac{1}{N^2 \alpha^r} \cdot \sigma_{\theta}^{r},
\end{equation}
\end{prop}
where $\sigma_{\theta}^{r} \triangleq  \frac{\sigma_w^2 }{1 + \frac{\sigma_w^2}{\alpha^r N^2 (\sigma^{r})^2}}$. \vspace{0.2cm}

We next establish a finite-sample bound on the error, given by the expected loss in the objective value at global round $R$ with respect to $F(\myVec{\theta}^*)$.

\begin{theorem}
    \label{Th1}
Suppose that each of the functions $f_i$ satisfies assumptions \emph{AS1,AS2,AS3}. Then, there exist weights $\{w_r\}$ and local step sizes $\eta_l \leq \frac{1}{8 \beta K (1+B^2)}$ such that the output (\ref{eq:final_output}) of the BAAF algorithm for any $R \geq \frac{8 \beta (1+B^2)}{\mu}$ satisfies:
\begin{align}
\label{eq:Th1}
    \nonumber &\mathbb{E}[F(\myVec{\bar{\theta}}^R)] -F(\myVec{\theta}^*) \\\nonumber &\leq 
    \left(\frac{3\sigma^2 (1+N)}{\mu RKN} + \frac{3d M^2 \sigma_{\theta}}{\mu RN^2P}\right) \cdot \log(\max\{\mu^2 R \frac{d_0}{c_1}\})   
    \\ \nonumber
    &+\frac{3\beta G^2}{\mu^2 R^2} \log^2(\max\{\mu^2 R \frac{d_0}{c_2}\} \\
    +&3\mu||\myVec{\hat{\myVec{\theta}}^0 - \theta^*} ||^2 \exp\left(- \frac{\mu R}{16 \beta (1+B^2)}\right),    
\end{align}
\end{theorem}
where  $\sigma_{\theta} \triangleq \displaystyle \max_{r \in [R], m \in [d]} \sigma_{\theta}^{r}$, $c_1 = \frac{\sigma^2 (1+N)}{KN} + \frac{d M^2 \sigma_{\theta}}{N^2P}$, $c_2 = \beta G^2$, $d_0 = ||\myVec{\hat{\myVec{\theta}}^0 - \theta^*} ||^2$. \\
The proof is given in Appendix \ref{sec:App_Th_1}.
Theorem \ref{Th1} gives a clean expression of how the four terms– the initial distance due to the error in the initial estimate, the variance $\sigma^2$ due to the stochastic gradients, the error $\sigma_\theta$ due to the noisy channel and the gradient dissimilarity parameter $G^2$ – characterize the error bound.

Comparing Theorem \ref{Th1} to the corresponding convergence rate of FedAvg derived in \cite[Theorem 1]{karimireddy2020scaffold}, which considered noise-free channels, we observe that BAAF achieves the same convergence rate, with an additional term which depends on the MSE induced by the channel noise. Larger MSEs make the convergence speed slower. By using the MMSE aggregation function that provides the minimum MSE, we can speed up the convergence rate. Additionally, by increasing $N$ this distortion term diminishes. 
When the MSE equals to $0$ (i.e., when $\sigma_\theta=0$), Theorem \ref{Th1} coincides with \cite[Theorem 1]{karimireddy2020scaffold}.
Additionally, comparing Theorem \ref{Th1} to the corresponding result in \cite{sery2021over}, which does not apply MMSE estimation to the channel output, we observe that the error term that depends on the channel noise in BAAF, $\sigma_\theta/P$, is smaller than the corresponding term $\sigma_w^2/P$ obtained in \cite{sery2021over}, which improves the convergence result.

Finally, the gradient dissimilarity parameter $G^2$ term in Th.~\ref{Th1} determines the decrease in the convergence rate as a result of the user-drift, which is a result of the non i.i.d nature of the users' data (i.e., the functions $f_i$ are distinct). In more details, let $\myVec{\theta}^*$ be the global optimum of $F(\myVec{\theta})$, and $\myVec{\theta}_i^*$ be the optimum of each user’s loss function $f_i(\myVec{\theta})$. In the case of heterogeneous data, it is quite likely that each of these $\myVec{\theta}_i^*$ is far away from the other, and from the global optimum $\myVec{\theta}^*$. The difference between $\frac{1}{N} \sum_{i=1}^N \myVec{\theta}_i^*$ (which is the server update) and the true optimum $\myVec{\theta}^*$ results in a user-drift. In the next section, we use the technique of control variates \cite{karimireddy2020scaffold,karimireddy2020mime} to mitigate this effect.
\color{black}

\section{The Controlled Bayesian Air Aggregation Federated learning (COBBAF) Algorithm}
\label{sec:CoBAAF}
We now extend the BAAF algorithm by implementing a variance reduction technique in the form of control variates in the local updates, thus reducing the negative effect of users-drift.

\subsection{Federated Averaging with Control Variates}
\label{ssec:SCAFFOLD}
The local updates of FedAvg (as in BAAF) leads to a drift in the update of each user resulting in slow and unstable convergence. We thus develop COBAAF using the technique of control variates for the local updates. As proposed in \cite{karimireddy2020scaffold, karimireddy2020mime}, along with the server model $\myVec{\theta}$, in each round $r$ we maintain a state for each user (user control variate $\myVec{c}_i^r \triangleq \myVec{g}_i(\myVec{\theta}^{r-1})$) and for the server (server control variate $\myVec{c}^r \triangleq \frac{1}{N} \sum_{i=1}^N \myVec{c}_i^r$). We first present the algorithm when executed over an error-free channel.
The iterative algorithm carries out multiple training rounds, each consisting of the following three phases:
\begin{enumerate}
    \item In global round $r$, starting from the shared model $\myVec{\theta}^r_{i,0}=\myVec{\theta}^{r-1}$, each user $i$ trains its local model using its local data set and the control variates over $K$ SGD local steps, namely, 
    \begin{equation}
        \label{eq:local_SCAFFOLD_Iteration}
    \begin{array}{c}
        \myVec{\theta}_{i,k}^r = \myVec{\theta}_{i,k-1}^r - \eta_l (\myVec{g}_i(\myVec{\theta}_{i,k-1}^r)-\myVec{c}_i^{r-1}+\myVec{c}^{r-1}),
    \end{array}
    \end{equation}
    \item Each user $i$ then updates the control variate: $\myVec{c}_i^r =  \myVec{g}_i(\myVec{\theta}^{r-1})$.
    \item Each user $i$ conveys its trained local model $\myVec{\theta}_{i,K}^r$ and the updated control variate $\myVec{c}_i^r$ to the central server, which averages them into a global model via $\myVec{\theta}^r = \frac{1}{N} \sum_{i=1}^N \myVec{\theta}_{i,K}^r
    $, and a global control variate $\myVec{c}^r = \frac{1}{N} \sum_{i=1}^N \myVec{c}_{i}^r$, and sends the new model and control to the users for another round.
\end{enumerate}

We next briefly explain the rationale behind the design of the algorithm. If communication is not a concern, then the ideal local update of user $i$ at global round $r$ would be 
$\myVec{\theta}_{i,k}^r= \myVec{\theta}_{i,k-1}^r - \frac{1}{N}\sum_{j=1}^N \myVec{g}_j(\myVec{\theta}_{i,k-1}^r)$. Such an update essentially computes an unbiased gradient of $F(\myVec{\theta})$ and hence becomes equivalent to running FedAvg in the i.i.d. case (which is known to have strong performance). Unfortunately, such an update requires communicating with all users for every update step. Instead, the procedure above uses control variates such that $\myVec{c}_i = \myVec{g}_i(\myVec{\theta}^{r-1})$ and $\myVec{c} = \frac{1}{N}\sum_{j=1}^N \myVec{c}_j$ for the whole round $r$. Since the gradient of $f_i$ is Lipschitz, it is expected that $\myVec{g}_i(\myVec{\theta}^{r-1}) \approx \myVec{g}_i(\myVec{\theta}_{i,k-1}^r)$ as long as the local updates are not too large and $\myVec{\theta}^{r-1} \approx \myVec{\theta}_{i,k-1}^r$. Then, the ideal update is estimated by:
$(\myVec{g}_i(\myVec{\theta}_{i,k-1}^r)-\myVec{g}_i(\myVec{\theta}^{r-1})+\frac{1}{N}\sum_{j=1}^N \myVec{g}_j(\myVec{\theta}^{r-1})) \approx \frac{1}{N}\sum_{j=1}^N \myVec{g}_j(\myVec{\theta}_{i,k-1}^r)$. Thus, the local updates do not drift and remain synchronized and converge under heterogeneous users.

We next address the aggregated scheme of COBAAF in the present of channel noise.

\subsection{Communication and Statistical Model}
\label{ssec:CoBBAF_System_Model}
In the iterative algorithm presented above the users are required to send additional information to the server, the controls $\myVec{c}_i \triangleq \myVec{g}_i(\myVec{\theta}^{r-1})$. 
However, we again observe that the server only needs to receive the average value $\frac{1}{N} \sum_{i=1}^N \myVec{c}_i$, where the individual control value $\{\myVec{c}_i\}$ is not required. Thus, each user precodes its control $\myVec{g}_i(\myVec{\theta}^{r-1})$ into the MAC channel input via $\sqrt{\beta^r} \myVec{g}_i(\myVec{\theta}^{r-1})$, where 
\begin{equation}
\label{eq:precoding_controls}
\beta^r \triangleq \frac{P}{\max_i \mathbb{E}[||\myVec{g}_i(\myVec{\theta}^{r-1})||^2]}.
\end{equation}
Then, together with  $\myVec{x}_i^r$ defined in (\ref{eq:channel_input}), the users transmit $\{\sqrt{\beta^r} \myVec{c}_i\}$ in an OTA fashion.
The channel outputs are given by
\begin{equation}
\label{eq:CoBAAF_channel_model}
  \Tilde{\myVec{y}}^r = \sum \limits_{i=1}^N \myVec{x}_i^r +\myVec{\Tilde{w}}^r \;, \quad \quad
 \Tilde{\myVec{z}}^r = \sum \limits_{i =1}^N \sqrt{\beta^r} \myVec{c}_i^r +\myVec{\tilde{n}}^r,
\end{equation}
where $\myVec{\Tilde{w}}^r,\myVec{\Tilde{n}}^r \sim \mathcal{N}(0,\sigma_w^2 \myVec{I}_d)$ are $d \times 1$ vectors of additive noise.
The server then sets:
\begin{equation}
\label{eq:CoBAAF_server_Pre}
  \Tilde{\myVec{\theta}}^r = \frac{\Tilde{\myVec{y}}^r}{N \sqrt{\alpha^r}} + \myVec{\theta}^{r-1}  \;, \quad \quad
 \Tilde{\myVec{c}}^r = \frac{\Tilde{\myVec{z}}^r}{N \sqrt{\beta^r}},
\end{equation}
which can be equivalently written as:
\begin{equation}
\label{eq:CoBAAF_equivalent_output}
  \Tilde{\myVec{\theta}}^r = \frac{1}{N} \sum \limits_{i=1}^N \myVec{\theta}_i^r +\myVec{w}^r \;, \quad  \quad 
 \Tilde{\myVec{c}}^r = \frac{1}{N} \sum \limits_{i =1}^N \myVec{c}_i^r +\myVec{n}^r,
\end{equation}
where $\myVec{w}^r \sim \mathcal{N}(0,\frac{\sigma_w^2}{N^2 \alpha^r} \myVec{I}_d)$ and $\myVec{n}^r \sim \mathcal{N}(0,\frac{\sigma_w^2}{N^2 \beta^r} \myVec{I}_d)$ are the equivalent noise terms. 

We also exploit the prior distribution of the gradients $\{\myVec{c}_i\}$, similarly as in Section \ref{ssec:Channel}, i.e., we assume that $\myVec{c}_i^r= [c_i^{r,1},\ldots c_i^{r,d}]$ are multivariate i.i.d Gaussian random variables with common mean $b_i^{r}$ and variance $(v_i^{r})^2$, 
which can be computed by performing offline simulations over a small data set.
where again it is assumed that $\myVec{c}_i^r$ and $\myVec{c}_n^r$ are uncorrelated for $i \neq n$.

\subsection{Algorithm Design}
\label{ssec:CoBAAF Algorithm}
We now present the steps of COBAAF. In each communication round, the server observes the \textit{noisy} sum of local models $\Tilde{\myVec{\theta}}^r$ and control variates $\Tilde{\myVec{c}}^r$, and first needs to estimate the average values $\myVec{\theta}^r = \frac{1}{N} \sum_{i=1}^N \myVec{\theta}_i^r$ and $\myVec{c}^r = \frac{1}{N} \sum_{i=1}^N \myVec{c}_i^r$. Thus it implements the conditional expectations:
\begin{equation}
\label{eq:CoBAAF_MSE_Solution}
        \hat{\theta}_{\text{MMSE}}^{r,m} = \mathbb{E}\left[\theta^{r,m} | \Tilde{\myVec{\theta}}^r \right] = \mu^r  + \frac{(\sigma^r)^2}{(\sigma^r)^2+ \frac{\sigma_w^2}{\alpha^r N^2}} (\Tilde{\theta}^{r,m} - \mu^r),
\end{equation}
\begin{equation}
\label{eq:CoBAAF_MSE_Solution_2}    
    \hat{c}_{\text{MMSE}}^{r,m} = \mathbb{E}\left[c^{r,m} | \Tilde{\myVec{c}}^r \right] = b^r  + \frac{(v^r)^2}{(v^r)^2+ \frac{\sigma_w^2}{\beta^r N^2}} (\Tilde{c}^{r,m} - b^r),
\end{equation}
where $ b^{r} \triangleq \frac{1}{N} \sum \limits_{i=1}^N b_i^{r}$ and $(v^{r})^2 \triangleq \frac{1}{N^2} \sum \limits_{i=1}^N (v_i^{r})^2$.
After computing the estimated sums in (\ref{eq:CoBAAF_MSE_Solution}),(\ref{eq:CoBAAF_MSE_Solution_2}), the server sends back $\myVec{\hat{\theta}}_{\text{MMSE}}^r,\myVec{\hat{c}}_{\text{MMSE}}^r$ to the users, and the local model is updated by: $\myVec{\theta}^{r+1}_{i,0}= \myVec{\hat{\theta}}_{\text{MMSE}}^r$. 
The local updated rule of user $i$ at round $r+1$ under COBAAF is then given by:
\begin{equation}
    \label{CoBAAF_real_local_update}
\myVec{\theta}_{i,k}^{r+1} = \myVec{\theta}_{i,k-1}^{r+1} - \eta_l (\myVec{g}_i(\myVec{\theta}_{i,k-1}^{r+1})-\myVec{c}_i^{r}+\myVec{\hat{c}}_{\text{MMSE}}^r), k=1,\ldots,K.    
\end{equation}
The final output of the algorithm is a weighted average for some positive weights (defined explicitly in Appendix \ref{sec:App_Th_2}) $\{w_r\}$ for $r \in \{1,\ldots,R+1\}$
\begin{equation}
\label{eq:CoBAAF_final_output}
    \myVec{\bar{ \theta}}^R = \frac{1}{\sum_r w_r} \sum_r w_r \myVec{\theta}^{r-1}.
\end{equation}
The pseudocode of COBAAF is given in Algorithm \ref{Algorithm2}.

\begin{algorithm}
\caption{COBAAF Algorithm} 
\label{Algorithm2}
  \begin{algorithmic}[1]
    \STATE  \textbf{server input} initial parameters $\myVec{\theta}, \myVec{c}$
    \STATE  \textbf{user input} local step size $\eta_l$, local control $\myVec{c}_i$, means $\mu_i^{r}, b_i^{r}$, variances $(\sigma_i^r)^2, (v_i^r)^2$
    \FOR{each communication round $r=1,\ldots,R$} 
        \STATE     communicate ($\myVec{\hat{\theta}}^{r-1}_{\text{MMSE}},\myVec{\hat{c}}^{r-1}_{\text{MMSE}} $) to all users
        \FOR{each user $i=1,\ldots,N$} 
            \STATE   initialize local parameters $\myVec{\theta_i}^r \leftarrow \myVec{\hat{\theta}}^{r-1}_{\text{MMSE}}$
            \FOR{each local step $k=1,\ldots,K$}
                \STATE compute a stochastic gradient $\myVec{g}_i(\myVec{\theta}_i^r)$ of $f_i$
                \STATE $\myVec{\theta_i^r} \leftarrow \myVec{\theta_i^r}-\eta_l (\myVec{g}_i(\myVec{\theta}_i^r) - \myVec{c_i^{r-1}} + \myVec{\hat{c}}^{r-1}_{\text{MMSE}})$
            \ENDFOR
            \STATE $\myVec{c}_i^r \leftarrow  \myVec{g}_i(\myVec{\theta}^r)$
            \STATE  communicate $\left(\myVec{\theta}_i^r,\myVec{c}_i^r, \mu_i^r,(\sigma_i^r)^2, m_i^r, (v_i^r)^2 \right)$
        \ENDFOR
        \STATE recover $\Tilde{\myVec{\theta}}^r, \Tilde{\myVec{c}}^r$ according to (\ref{eq:CoBAAF_server_Pre})
        \STATE Estimate $\hat{\theta}^{r,m}_{\text{MMSE}} = \mu^r  + \frac{(\sigma^r)^2}{(\sigma^r)^2+ \frac{\sigma_w^2}{\alpha^r N^2}} (\Tilde{\theta}^{r,m} - \mu^r)$    
        \STATE Estimate $\hat{c}^{r,m}_{\text{MMSE}} = b^r  + \frac{(v^r)^2}{(v^r)^2+ \frac{\sigma_w^2}{\beta^r N^2}} (\Tilde{c}^{r,m} - b^r)$        
    \ENDFOR
  \end{algorithmic}
\end{algorithm}

\subsection{Theoretical Analysis}
\label{ssec:Theoretical_CoBAAF}
We next establish a finite-sample bound on the error of the COBAAF algorithm, given by the expected loss in the objective value at global round $R$ with respect to $F(\myVec{\theta}^*)$.

\begin{theorem}
\label{Th2}
Suppose that each of the functions $f_i$ satisfies assumptions \emph{AS1,AS2}.
Then, there exist weights $\{w_r\}$ and local step sizes $\eta_l \leq \frac{1}{8 \beta K}$ such that the output (\ref{eq:final_output}) of the COBAAF algorithm for any $R \geq \frac{8 \beta}{\mu}$ satisfies:
\begin{align}
\label{eq:Th2}
    \nonumber &\mathbb{E}[F(\myVec{\bar{\theta}}^R)] -F(\myVec{\theta}^*) \\\nonumber \leq &\left(\frac{2\sigma^2 (1+N)}{\mu RKN} + \frac{2d M^2 \sigma_{\theta}}{\mu R K N^2 P}   
    +\frac{2d M^2 \sigma_c (1+N K)}{\mu R K N^2 P}\right) \\ \nonumber &\cdot \log(\max \{1, \mu^2 R d_0/c_3 \})\\
    +&2\mu||\myVec{\hat{\myVec{\theta}}^0 - \theta^*} ||^2 \exp\left(- \frac{\mu R}{16 \beta}\right),
\end{align}
where  $\sigma_{c}^r=\max_{r\in [R], m\in [d]} \sigma_{c}^{r}$, $\sigma_{c}^{r} \triangleq  \frac{\sigma_w^2 }{1 + \frac{\sigma_w^2}{\beta^r \sum_{i=1}^N (v_i^{r})^2}}$, $c_3 = \frac{\sigma^2 (1+N)}{KN} + \frac{d M^2 \sigma_{\theta}}{K N^2 P}   
    +\frac{d M^2 \sigma_c (1+N K)}{K N^2 P}$.
\end{theorem}
The proof is given in Appendix \ref{sec:App_Th_2}.

Similarly to Theorem \ref{Th1}, Theorem \ref{Th2} gives a clean expression of how the four terms– the initial distance due to the error in the initial estimate, the variance $\sigma^2$ due to the stochastic gradients, and the errors $\sigma_\theta$, $\sigma_c$ due to the noisy channel – characterize the error bound.
Comparing Theorem \ref{Th2} to the corresponding result in \cite{karimireddy2020scaffold}, which considered noise-free channels, we observe that COBAAF achieves the same convergence rate, with an additional terms which depend on the MSE induced by the channel noise and fading. Larger MSEs make the convergence speed slower. By using the MMSE aggregation function that provides the minimum MSE, we can speed up the convergence rate. Additionally, by increasing $N$ this distortion term diminishes. 
When the MSE equals $0$, Theorem \ref{Th2} coincides with \cite[Theorem 3]{karimireddy2020scaffold}.
Additionally, comparing Theorem \ref{Th2} to the corresponding result in Theorem \ref{Th1} and in \cite{sery2021over}, which considered the same setup but without addressing the statistical heterogeneity, we observe that Theorem \ref{Th1} and Theorem $1$ in \cite{sery2021over} have an additional component which depends on $G^2$ and $\Gamma\triangleq F^*- \frac{1}{N}\sum_{n=1}^N f_n^*$, respectively, that encapsulates the degree of heterogeneity. The use of control variates in COBAAF mitigates the negative effect of this term on the convergence.

Theorem \ref{Th2} implies an important trade-off between communication and heterogeneity: To reduce heterogeneity, the users are required to send the additional parameters $\myVec{c}_i$, which results in the error term $\sigma_c$. However, the use of the control variates mitigates the negative effect of the dissimilarity term $G^2$ on the convergence. Finally, note that due to the OTA transmission of the COBAAF algorithm, the bandwidth requirement is independent of the number of users $N$, allowing the simultaneous participation of a large number of users without limiting the throughput of each user. Practically, this transmission scheme requires coherent transmissions between users \cite{gafni2022federated}.
\color{black}

\subsection{Extension to Fading Channels}
\label{ssec:fading_channels}
In the previous subsections we focused on FL over noisy MACs, where $\tilde{h}_i^r=1$ for all $r, i$. Here, we extend COBAAF to handle transmissions over fading MAC with general fading values, while preserving its proven convergence (the extension of BAAF to fading MACs can be done similarly).
In fading MACs, the signal transmitted by each user undergoes a fading coefficient denoted by $h_i^r e^{j \phi_i^r}$ as defined in (\ref{eq:channel_model}) for the local model updates, and $\rho_i^r e^{j \varphi_i^r}$ for the control variates. In the extension here, we assume that the participating entities have channel state information (CSI), i.e., knowledge of the fading coefficients, as commonly assumed in wireless communications and particularly OTA FL (see e.g., \cite{cohen2010time, zhu2020one, sery2021over} and subsequent studies). Such knowledge can be obtained by letting the users sense their channels, estimating the channel from pilot signals, or having the access point/server periodically estimate these coefficients and convey them to the users.

Following the scheme proposed in \cite{sery2021over}, each user utilizes its CSI to cancel the fading effect by amplifying the signal by its inverse channel coefficient. However, weak channels might cause an arbitrarily high amplification, possibly violating the transmission power constraint (\ref{eq:energy_constraint}). Therefore,  thresholds $h_{\min}, \rho_{\min}$ are set, and users observing fading coefficients of a lesser magnitude than $h_{\min}$ (for the model updates) or $\rho_{\min}$ (for the control variates) do not transmit in that communication round.
As channels typically attenuate their signals, it holds that $h_{\min} < 1, \rho_{\min} < 1$. Under this extension of COBAAF, (\ref{eq:channel_input}) becomes 
\begin{equation}
\label{eq:fading_input}
        \displaystyle \myVec{x}_i^r \!=\! 
    \begin{cases}
    \frac{\sqrt{\alpha^r} h_{\min}}{h_i^r} e^{-j\phi_i^r} (\myVec{\theta}_i^r - \myVec{\theta}_{i,0}^r) , & \text{if} \; h_i^r > h_{\min},    \\ 
    0, & \text{if} \; h_i^r \leq h_{\min}.
    \end{cases} 
    \end{equation}
Note that the energy constraint (\ref{eq:energy_constraint}) is preserved.
The control variates $c_i$ are transmitted in a similar way.

To formulate the server aggregation, we let $\mathcal{S}^r \subset \mathcal{N}$ be the set of user indices whose corresponding channel at round $r$ satisfies $h_i^r > h_{\min}$, $\rho_i^r > \rho_{\min}$. As the server has CSI, it knows $\mathcal{S}^r$, and can thus recover the aggregated model $\myVec{\theta}^r$ and control $\myVec{c}^r$ in a similar manner as in (\ref{eq:CoBAAF_server_Pre}), via
\begin{equation}
\label{eq:fading_CoBAAF_server_Pre}
  \Tilde{\myVec{\theta}}^r = \frac{\Tilde{\myVec{y}}^r}{|\mathcal{S}^r| \sqrt{\alpha^r} h_{\min}} + \myVec{\theta}^{r-1}  \; \quad \quad
 \Tilde{\myVec{c}}^r = \frac{\Tilde{\myVec{z}}^r}{|\mathcal{S}^r| \sqrt{\beta^r} \rho_{\min}},
\end{equation}
which can be equivalently written as:
\begin{align}
\label{eq:fading_CoBAAF_equivalent_output}
  \nonumber &\Tilde{\myVec{\theta}}^r = \frac{1}{|\mathcal{S}^r|} \sum \limits_{i\in \mathcal{S}^r} \myVec{\theta}_i^r +\frac{N}{|\mathcal{S}^r| h_{\min}}\myVec{w}^r, \\ &\Tilde{\myVec{c}}^r = \frac{1}{|\mathcal{S}^r|} \sum \limits_{i\in \mathcal{S}^r} \myVec{c}_i^r + \frac{N}{|\mathcal{S}^r| \rho_{\min}}\myVec{n}^r.
\end{align}
As in Section \ref{ssec:CoBAAF Algorithm}, given $\tilde{\myVec{\theta}}^r,\Tilde{\myVec{c}}^r$, the server estimates the partial sums $\frac{1}{|\mathcal{S}^r|} \sum_{i\in \mathcal{S}^r} \myVec{\theta}_i^r$, $\frac{1}{|\mathcal{S}^r|} \sum_{i\in \mathcal{S}^r} \myVec{c}_i^r$, using the MMSE estimator, and sends the estimators $\hat{\myVec{\theta}}^r, \hat{\myVec{c}}^r$ to the users for the next round.

Comparing (\ref{eq:fading_CoBAAF_equivalent_output}) to the corresponding equivalent formulation in (\ref{eq:CoBAAF_equivalent_output}), we note that the proposed extension of COBAAF
results in two main differences from the fading-free scenario: First, the presence of fading is translated into an increase in the noise power, encapsulated in the constants $\frac{N}{|\mathcal{S}^r| h_{\min}} > 1$, $\frac{N}{|\mathcal{S}^r| \rho_{\min}} > 1$. Second, fewer models are aggregated in each round as $\mathcal{S}^r \leq N$. 
The set of participating users $\mathcal{S}^r$ depends on the distribution of the fading coefficients. Thus, in order to analytically
characterize how the convergence is affected by fading compared to the scenario analyzed in Subsection \ref{ssec:Theoretical_CoBAAF}, we
introduce the following assumption: \\ 
\textit{AS4}: At each communication round, the participating users set $\mathcal{S}^r$ contains $S \leq N$ users and is uniformly distributed over all the subsets of $\mathcal{N}$ of cardinality $S$.
Note that Assumption \textit{AS4} can be imposed by a simple distributed mechanism using an opportunistic carrier sensing as described in \cite{sery2021over}.

We point out that the partial aggregation scheme induces some 'lag' in our updates of the control variates $\myVec{c}_i$ since only a small subset of them are updated at each round. That is, the global control $\myVec{c}^r = \frac{1}{S} \sum_{i \in \mathcal{S}^r} \myVec{c}_i^r$ also drifts from the "desired" average $\sum_{i \in \mathcal{N}} \myVec{c}_i^r$. We address this point in the proof of Theorem \ref{Th3}.

Next, we characterize the convergence of the instantaneous global model, as stated in the following theorem:

\begin{theorem}
\label{Th3}
Suppose that each of the functions $f_i$ satisfies assumptions \emph{AS1, AS2, AS4}. Then, there exist weights $\{w_r\}$ and local step sizes $\eta_l \leq \min \left(\frac{1}{81 \beta K}, \frac{S}{15 \mu N K}  \right) $ such that the output (\ref{eq:final_output}) of the COBAAF algorithm for any $R \geq \max \left(\frac{162 \beta}{\mu},\frac{30N}{S}   \right)$ satisfies:
\begin{align}
\label{eq:Th3}
    \nonumber &\mathbb{E}[F(\myVec{\bar{\theta}}^R)] -F(\myVec{\theta}^*) \\
    \nonumber &\leq \left(\frac{\sigma^2 (1+S)}{\mu RKS} + \frac{d M^2 \sigma_{\theta}}{\mu R K S^2 h_{\min}^2 P}   
    +\frac{d M^2 \tilde{\sigma}_c (1+S K)}{\mu R K S^2 h_{\min}^2 P} \right)  \\
    \nonumber & \cdot \log(\{1, \mu^2 R \tilde{D^2}/c_4 \}) \\
    +&\mu \tilde{D}^2  \exp\left(- \min \left\{\frac{S}{30N}, \frac{\mu}{162 \beta} \right\} R\right),
\end{align}
where $\tilde{D^2} \triangleq (||\myVec{\hat{\myVec{\theta}}^0 - \theta^*} ||^2 + \frac{1}{2 S \beta^2} \sum_{i=1}^N ||c_i^0 - \nabla f_i(\theta^*)||^2)$, $c_4 = \frac{\sigma^2 (1+S)}{KS} + \frac{d M^2 \sigma_{\theta}}{K S^2 h_{\min}^2 P}   
    +\frac{d M^2 \tilde{\sigma}_c (1+S K)}{K S^2 h_{\min}^2 P}$.
\end{theorem}
The proof is given in Appendix \ref{sec:App_Th_3}.

Comparing Theorem \ref{Th3} to the corresponding convergence bound with fading-free channels in Theorem \ref{Th3} reveals that the extension of COBAAF allows the trained model to maintain its asymptotic convergence rate of $O(\frac{1}{R})$ in the presence of fading channel conditions as well. However, the aforementioned differences in the equivalent global model due to fading are translated here into additive terms increasing the bound on the distance between the expected instantaneous objective $\mathbb{E}[F(\myVec{\bar{\theta}}^R)]$ and its desired optimal value.
In particular, the increased equivalent noise results in the terms $\tilde{\sigma}_c, \tilde{\sigma}_\theta$ being larger than the corresponding errors $\sigma_c, \sigma_{\theta}$ in (\ref{eq:Th2}) due to the increased equivalent noise-to-signal ratio which stems
from the scaling by $h_{\min}$ and $\rho_{\min}$ at the precoder and the corresponding aggregation at the server side.
Moreover, due to the partial participation of users in the global rounds, we also need to keep track how far our control variate is from its value at the optimum as discussed above. This is encapsulated in the term $\tilde{D}^2$, which is different (and larger) than the corresponding $D^2$ term in (\ref{eq:Th2}). 
In particular, note that the $\tilde{D}^2$ defined above involves an additional term $\frac{1}{2 S \beta^2} \sum_{i=1}^N ||c_i^0 - \nabla f_i(\theta^*)||^2)$ due to using a variance reduction method \cite{johnson2013accelerating}. Practically, we will use a warm-start strategy to set $\myVec{c}_i^0$ and in the first $N/S$ rounds, we compute $\myVec{c}_i^0 = \myVec{g}_i(\myVec{\theta}^0)$ over a batch size of size $K$. Then, using smoothness of $f_i$, we can bound this additional term as
\begin{align}
\nonumber
    &\frac{1}{2 S \beta^2} \sum_{i=1}^N ||c_i^0 - \nabla f_i(\theta^*)||^2) \leq \frac{N}{S \beta}(f(\myVec{\theta}^0) - f^*) + \frac{N \sigma^2}{KS\beta^2} \\ \nonumber
    & \leq \frac{ND^2}{S} + \frac{N \sigma^2}{KS\beta^2}.
\end{align}

\color{black}
\section{Numerical Analysis}
\label{sec:Simulations}

We now provide simulation results to demonstrate the efficiency of BAAF and COBAAF, in both regression and classification tasks.
We evaluate BAAF and COBAAF on real data sets. In order to better characterize statistical heterogeneity and study its effect on convergence, we also evaluate them on a set of synthetic data, which allows for more precise manipulation of statistical heterogeneity.

\subsection{Linear Regression on Synthetic Data}
\begin{figure}[htbp]
\centering \epsfig{file=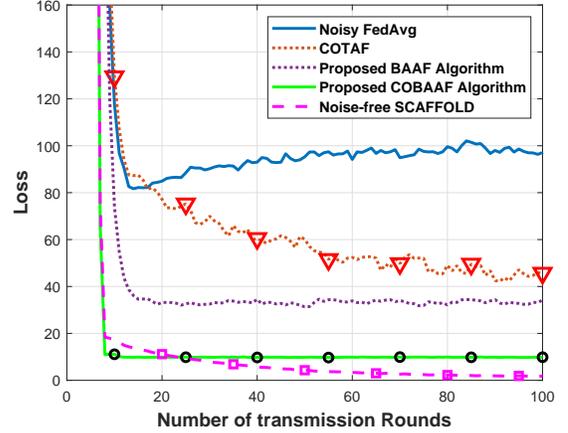, width=0.45\textwidth}
\vspace{-0.2cm}
\caption{Simulation results of a linear regression model. The loss as a function of the number of transmission rounds.}
\label{fig:fig1}
\end{figure}

We consider a linear regression model with a convex local loss function, i.e., $f_i(\myVec{\theta}) = || \myVec{A}_i^T \myVec{\theta} - \myVec{B}_i||_2^2$, where $\myVec{A}_i$ is the training input and $\myVec{B}_i$ is the corresponding level,  with $[D_i,d] = [100,10]$. In Fig.~\ref{fig:fig1}, we present the performance evaluation when the number of users is $N=20$, the number of local steps is $K=10$ and the local step size is $10^{-2}$. 
To embrace the data heterogeneity, for each user $i$ we model $\myVec{A}_i \sim \mathcal{N}(a_i, \myVec{I}), a_i \sim \mathcal{N}(1,\alpha), \myVec{\theta}_i \sim \mathcal{N}(b_i, \emph{I}), b_i \sim \mathcal{N}(-4, \beta)$, where $\emph{I}$ is the identity matrix. Therefore, $\alpha$ controls how much the local data at each user differs from that of other users, and $\beta$ controls how much local models differ from each other. We simulate MACs with signal-to-noise ratios (SNRs) of 10dB and without fading.
The precoding coefficients $\alpha^r,\beta^r$ are computed via (\ref{eq:precoder}),(\ref{eq:precoding_controls}), respectively, using numerical averaging, i.e., we carried out offline simulations of local SGD without noise and with $20\%$ of the data samples, and computed the averaged norm of the resulting model updates.
We numerically evaluate the gap from the achieved expected objective and the loss-minimizing one, i.e.,
$\mathbb{E}[F(\hat{\myVec{\theta}}^r)] - F(\myVec{\theta}^*)$.
Using this performance measure, we compare BAAF and COBAAF to the following FL methods: (i) Non-precoded OTA FL (denoted in Fig.~\ref{fig:fig1} as "Noisy FedAvg"), where every user transmits its model updates over the MAC without time varying precoding (\ref{eq:precoder}) and with a constant amplification as in \cite{sery2021over}, i.e., $\myVec{x}_i^r = P(\myVec{\theta}_i^r - \myVec{\theta}_{i,0}^r)$; (ii) COTAF \cite{sery2021over}, in which a time varying precoding is implemented similarly as in (\ref{eq:precoder}), but without the use of control variates and without Bayesian aggregation at the server side; and (iii) SCAFFOLD \cite{karimireddy2020scaffold} carried out over ideal orthogonal noiseless channels.
The stochastic expectation is evaluated by averaging over $100$ Monte Carlo trials, where in each trial the initial $\myVec{\theta}_0$ is randomized from zero-mean Gaussian distribution with covariance $\myVec{I}_d$.

It can be seen in Fig.~\ref{fig:fig1} that BAAF (purple curve) outperforms both FedAvg carried over noisy channels (blue curve), and COTAF (red curve), indicating the gain that we achieve by using the proposed precoding scheme and Bayesian aggregation method at the server side. Note that COTAF uses precoding but does not implement Bayesian aggregation, thus the gap between BAAF and COTAF indicates the performance gain as a result of the Bayesian aggregation scheme.    
We also observe that COBAAF outperforms BAAF and COTAF. This implies that COBAAF indeed reduces the variance induced by heterogeneity using the control variates compared to BAAF and COTAF that do not use this technique.  Note that BAAF uses precoding and implements Bayesian aggregation but does not use control variates in the local updates, thus the gap between COBAAF and BAAF indicates the performance gain as a result of the control variates. 
Finally, we also observe that COBAAF (green curve) achieves performances within a minor gap from that of SCAFFOLD \cite{karimireddy2020scaffold} carried out over ideal orthogonal noiseless channels (pink curve). That indicates that COBAAF is able to mitigate the noise effect. 
It is emphasized that when using orthogonal transmissions,
as implicitly assumed when using the conventional SCAFFOLD scheme, increasing the number of users implies that the channel resources must be shared among a larger number of users, hence the throughput of each user decreases. However, in OTA FL the throughput is independent of the number of users. The results presented in Fig.~\ref{fig:fig1} demonstrate the benefits of COBAAF, as an OTA FL scheme which accounts for both the convergence properties of FL algorithms as well as the unique characteristics of wireless communication channels.

\begin{figure}[htbp]
\centering \epsfig{file=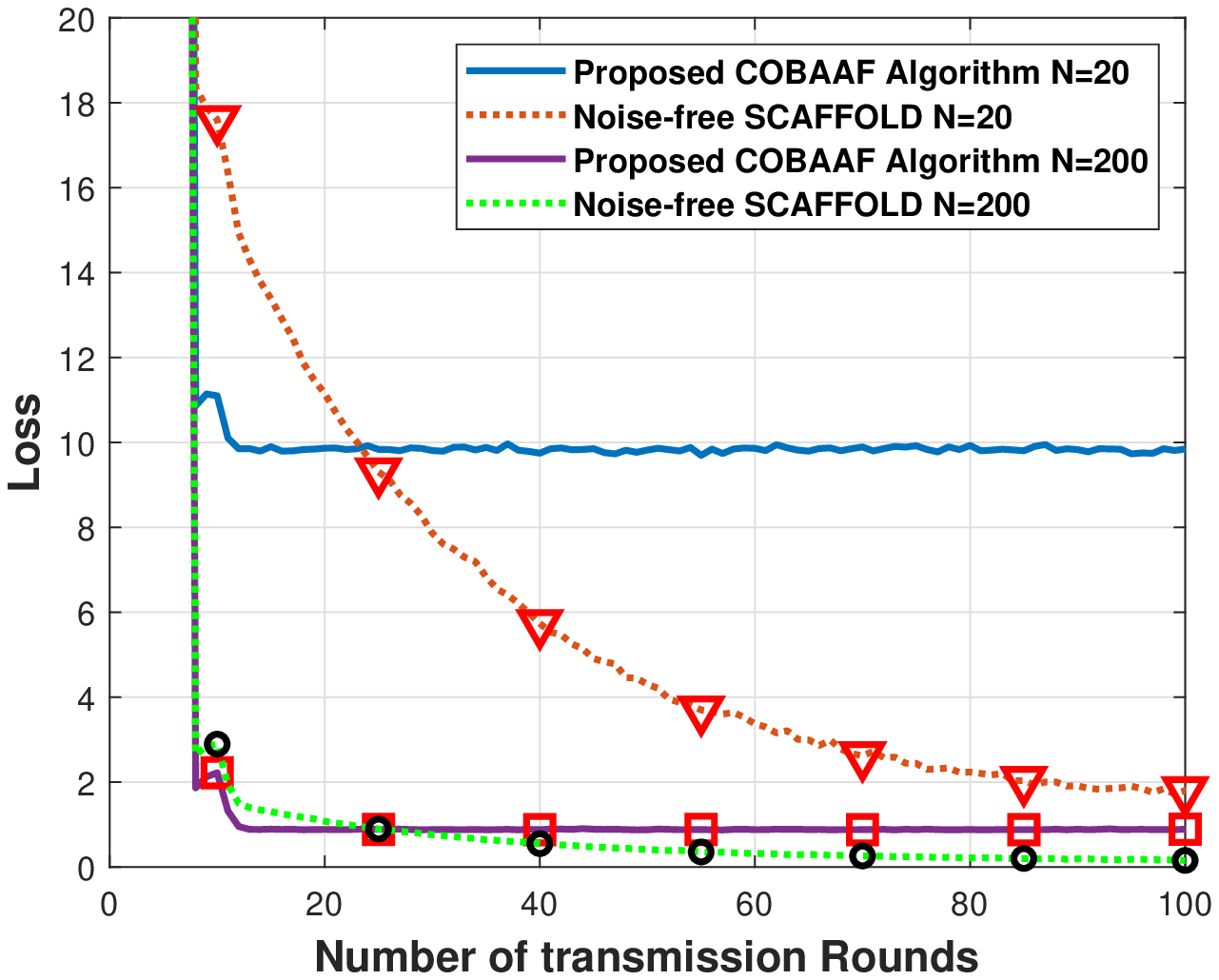, width=0.45\textwidth}
\vspace{-0.2cm}
\caption{Simulation results of a linear regression model. The loss as a function of the number of transmission rounds, where the number of users increases.}
\label{fig:figN}
\end{figure}

\color{black}
Next, we repeat the simulation study of Fig.~\ref{fig:fig1} while increasing the number of users to be $N=200$ in Fig.~\ref{fig:figN}. The results demonstrate the dependence of COBAAF on the number of users $N$. Since the MSE resulting from the Bayesian aggregation decreases when $N$ increases, increasing the number of users $N$ improves the performance of COBAAF. In particular, COBAAF almost reaches the performance of noise-free SCAFFOLD.   

\begin{figure}[htbp]
\centering \epsfig{file=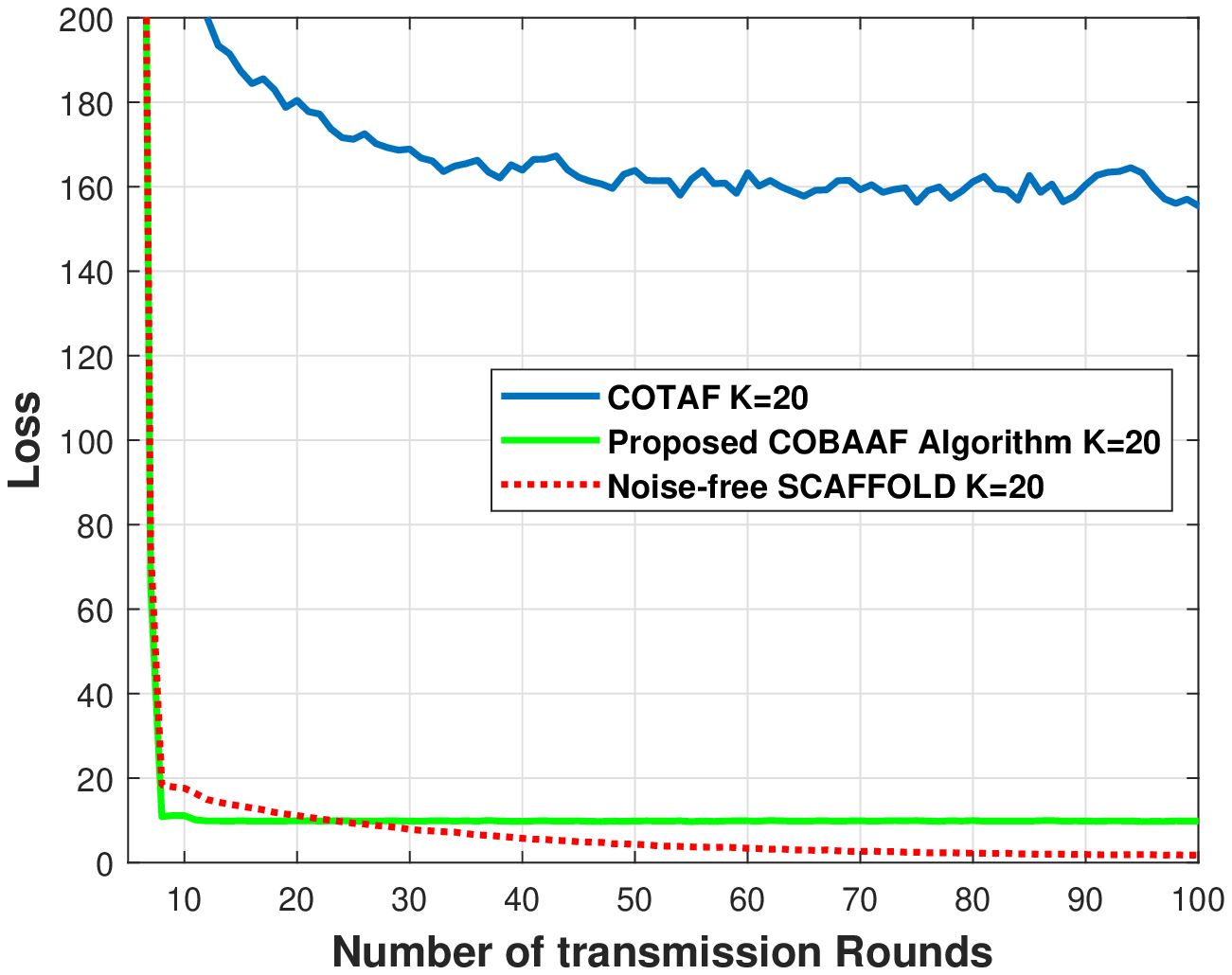, width=0.45\textwidth}
\vspace{-0.2cm}
\caption{Simulation results of a linear regression model. The loss as a function of the number of transmission rounds, where the number of local steps increases.}
\label{fig:figK}
\end{figure}

Finally, we repeat the simulation study of Fig.~\ref{fig:fig1} while increasing the number of local steps to be $K=20$ in Fig.~\ref{fig:figK}
Comparing Fig.~\ref{fig:figK} to Fig.~\ref{fig:fig1} reveals that increasing the SGD steps $K$ can improve the performance of COBAAF as the channel noise is induced less frequently. We also observe that the gap between COBAAF to COTAF increases compared to Fig.~\ref{fig:fig1}, since performing more local steps in heterogeneous systems increases the users drift in the COTAF algorithm. The control variates $c,c_i$ in COBAAF allows to utilize the local steps to improve the performance.  

\subsection{Classification on MNIST Data set}
\begin{figure}[htbp]
\centering \epsfig{file=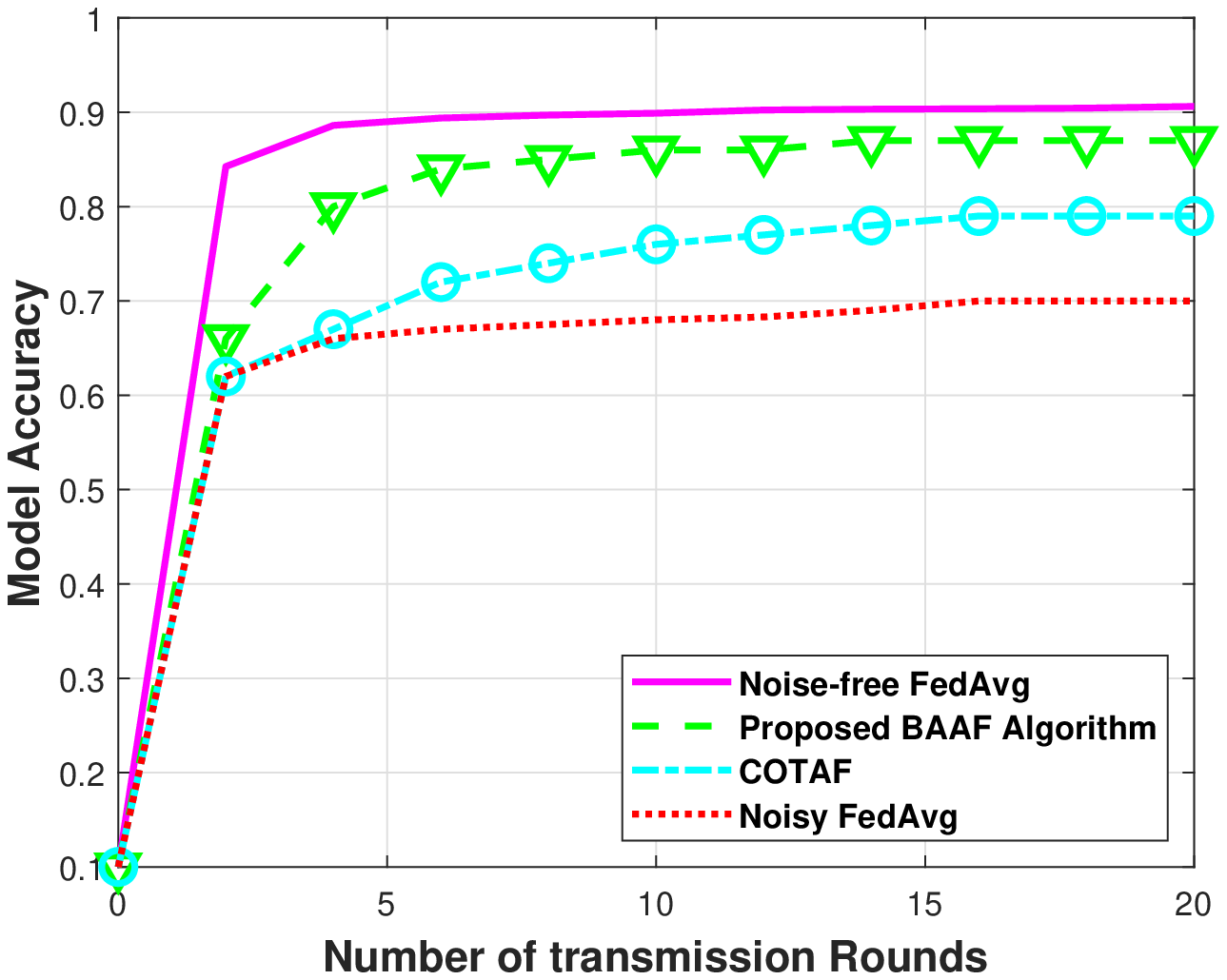, width=0.45\textwidth}
\vspace{-0.2cm}
\caption{Simulation results of a logistic regression model on the MNIST data set in the homogeneous setting. The accuracy as a function of the number of transmission rounds.}
\label{fig:iid}
\end{figure}

\begin{figure}[htbp]
\centering \epsfig{file=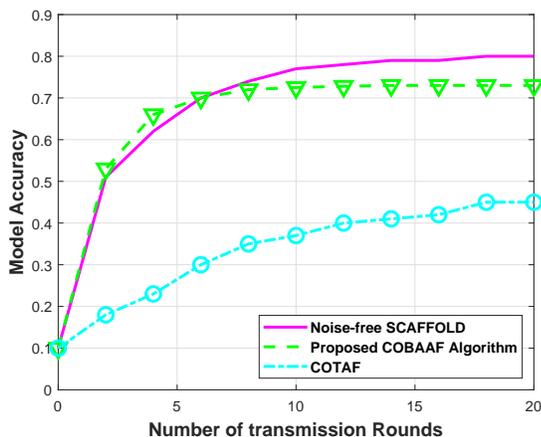, width=0.45\textwidth} \vspace{-0.2cm}
\caption{Simulation results of a logistic regression model on the MNIST data set in the heterogeneous setting. The accuracy as a function of the number of transmission rounds.}
\label{fig:non_iid}
\end{figure}

Next, we consider an image classification problem, based on the MNIST data set \cite{lecun1998gradient}, which contains train and test images from ten different digits. Our real data set experiments run logistic regression (convex) model as in \cite{karimireddy2020scaffold}.
We study two ways of partitioning the MNIST data over users: balanced partitioning, where the data is shuffled, and then partitioned into 10 users each receiving 600 examples, and imbalanced partitioning, where approximately $20\%$ of the training data of each user is associated with a single label, which differs among the different users. Consequently, each user holds more images from a unique class, resulting in heterogeneity between the users. The model accuracy versus the transmission round achieved for the considered FL schemes is depicted in Figs.~\ref{fig:iid} and \ref{fig:non_iid} for the homogeneous case and the heterogeneous case, respectively.
We evaluate the performances when the number of local steps is $K=5$, the number of users is $N=10$, the SNR is $10$dB without fading, and the learning rate is $10^{-2}$.

Observing Fig.~\ref{fig:iid}, we note that in the homogeneous setting, BAAF (green curve) outperforms both FedAvg over noisy channels (red curve), and COTAF (cyan curve), indicating the gain that we achieve also in this classification task by using the proposed precoding scheme and Bayesian aggregation method at the server side. The gap between BAAF and COTAF indicates the performance gain as a result of the Bayesian aggregation scheme. Finally, we also observe that BAAF achieves performances within a minor gap from that of FedAvg (purple curve) carried over ideal orthogonal noiseless channels (which has great performances in the homogeneous case). 

Finally, in Fig.~\ref{fig:non_iid} we compared the performances of the algorithms in the MNIST classification task in the heterogeneous setting over noisy channels. We observe that the proposed CoBAAF algorithm outperforms COTAF, with a larger gap than in the homogeneous case. This is due to the ability of CoBAAF to handle heterogeneous data sets. By employing precoding and Bayesian aggregation, COBBAF is able to almost achieve the performance of SCAFFOLD over a noiseless channel.

\vspace{-0.3cm}
\section{Conclusions}
\label{sec:conclusions}
In this paper, we developed two OTA FL algorithms for combating noise and heterogeneity. Our proposed BAAF algorithm is based on a rigorous Bayesian approach for signal aggregation to mitigate the noise and fading effects induced by the channel. To handle heterogeneity, we developed COBAAF, which extends BAAF and uses a control variates technique. We rigorously prove that the BAAF and COBAAF converge to the loss minimizing model with the same asymptotic convergence rate of FedAvg \cite{mcmahan2017communication} and SCAFFOLD \cite{karimireddy2020scaffold} (respectively) without communication constraints. The strong performances of COBAAF are achieved by combining the transmission and Bayesian aggregation scheme, with proper updates at the user side. Numerical results demonstrate strong performance of the method.

\appendices{}

\vspace{-0.2cm}
\section{Technical Lemmas}
\label{ssec:Technical_Lemmas}
In this section we cover some technical lemmas which are useful for computations later on.
The lemma below is useful to unroll recursions and derive convergence rates.
\begin{lemma} (convergence rate)
\label{lemma:convergence rate}
For every non-negative sequence $\{d_{r-1}\}$ and any parameters $\mu>0, \eta_{\max} \in (0,\frac{1}{\mu}], c\geq 0, R\geq \frac{1}{2\eta_{\max} \mu} $, there exists a constant step-size $\eta \leq \eta_{\max}$ and weights $w_r := (1-\mu \eta)^{1-r}$ such that for $W_R := \sum_{r=1}^{R+1} w_r$,
\begin{align}
\nonumber &\frac{1}{W_R} \sum \limits_{r=1}^{R+1} \left(\frac{w_r}{\eta}(1-\mu \eta)d_{r-1} - \frac{w_r}{\eta} d_r +c \eta w_r \right) \\\nonumber
&= O\left(\mu d_0 \exp (-\mu \eta_{\max}R) + \frac{c}{\mu R} \right).
\end{align}
\end{lemma}

\textit{Proof}: see Lemma $1$ in \cite{karimireddy2020scaffold}.

\begin{lemma} (Separating mean and variance)
\label{lemma:separating}
Let $\{\Xi_1, \ldots \Xi_\tau \}$ be $\tau$ random variables in $\mathbb{R}^d$ which are not necessarily
independent. First suppose that their mean is $\mathbb{E}[\Xi_i] = \xi_i$ and variance is bounded as $\mathbb{E}[||\Xi_i - \xi_i||^2] \leq \sigma^2 $. Then, the following holds:
\begin{equation*}
    \mathbb{E}[||\sum \limits_{i=1}^{\tau} \Xi_i ||^2 ] \leq ||\sum \limits_{i=1}^{\tau} \xi_i ||^2 + \tau^2 \sigma^2.
\end{equation*}
Now instead suppose that their conditional mean is $\mathbb{E}[\Xi_i|\Xi_{i-1},\ldots \Xi_1 ] = \xi_i$, i.e., the variables $\{\Xi_i - \xi_i \}$ form a martingale
difference sequence, and the variance is bounded by $\mathbb{E}[||\Xi_i - \xi_i||^2] \leq \sigma^2 $ as before. Then we can show the tighter bound 
\begin{equation*}
    \mathbb{E}[||\sum \limits_{i=1}^{\tau} \Xi_i ||^2 ] \leq 2||\sum \limits_{i=1}^{\tau} \xi_i ||^2 + 2\tau \sigma^2.
\end{equation*}
\end{lemma}

\textit{Proof}: see Lemma $4$ in \cite{karimireddy2020scaffold}.

\section{Proof of Eq.~(\ref{eq:MSE_Solution}) and Proposition \ref{lemma:MSE}}
\label{sec:App_Lemma_1_2}
\textit{Proof of Eq.~(\ref{eq:MSE_Solution})}: Since the elements of $\myVec{\theta}_i^r$ are independent, we have: $\mathbb{E}[\theta^{r,m} | \Tilde{\myVec{\theta}}^r] = \mathbb{E}[\theta^{r,m} | \Tilde{\theta}^{r,m}]$. We also observe that $\theta^{r,m}$ and $ \Tilde{s}^{r,m}$ are jointly Gaussian, therefore we have: 
\begin{align}
    \label{eq:app1}
\nonumber &\hat{\theta}^{r,m}_{\text{MMSE}} = \mathbb{E}[\theta^{r,m} | \Tilde{\theta}^{r,m}] \\ &= \mathbb{E}[\theta^{r,m}] + \frac{\text{cov}(\theta^{r,m},\Tilde{\theta}^{r,m})}{\text{var}(\Tilde{\theta}^{r,m})} (\Tilde{\theta}^{r,m} - \mathbb{E}[\Tilde{\theta}^{r,m}]). 
\end{align}
We next compute the required components: \vspace{0.2cm} \\
$\mathbb{E}[\theta^{r,m}] =  \mathbb{E}[\frac{1}{N} \sum_{i=1}^N \theta^{r,m}_i] = \frac{1}{N}\sum_{i =1}^N \mu_i^{r} = \frac{1}{N} \mu^{r}$ \vspace{0.2cm} \\
$\mathbb{E}[\Tilde{\theta}^{r,m}] =  \mathbb{E}[\frac{1}{N} \sum_{i=1}^N \theta^{r,m}_i + w^{r,m}] =  \frac{1}{N} \mu^r$ \vspace{0.2cm} \\
$\text{cov}(\theta^{r,m},\Tilde{\theta}^{r,m}) \vspace{0.2cm} \\ = \text{cov}(\frac{1}{N}\sum_{i =1}^N \theta^{r,m}_i ,\frac{1}{N} \sum_{l =1}^N \theta^{r,m}_l + w^{r,m}) = \frac{1}{N^2} \text{var}(\theta^{r,m}) \vspace{0.2cm} \\ = \frac{1}{N^2} (\sigma^{r})^2$, \vspace{0.2cm} \\
where the second inequality in the covariance evaluation holds since $\theta^{r,m}_i$ and $\theta^{r,m}_l$ are independent for $i \neq l$. Finally, \vspace{0.2cm} \\
$\text{var}(\Tilde{\theta}^{r,m}) = \text{var}(\frac{1}{N} \sum_{i=1}^N \theta^{r,m}_i + w^{r,m}) =  \frac{1}{N^2} (\sigma^{r})^2 + \sigma_w^2$. \vspace{0.2cm} \\
By plugging the above components in (\ref{eq:app1}), we obtain (\ref{eq:MSE_Solution}).

\textit{Proof of Proposition \ref{lemma:MSE}}:
By utilizing again the fact that the elements of $\myVec{\theta}_i^r$ are independent and $\theta^{r,m}$ and $\Tilde{\theta}^{r,m}$ are jointly Gaussian, we have: 
\begin{equation}
    \label{eq:app2}
\text{MSE}(\hat{\theta}^{r,m}_{\text{MMSE}}) = \text{var}(\theta^{r,m}) - \frac{(\text{cov}(\theta^{r,m},\Tilde{\theta}^{r,m}))^2}{\text{var}(\Tilde{\theta}^{r,m})}. \end{equation}
Since $\text{var}(\theta^{r,m}) = \sum_{i=1}^N (\sigma_i^{r})^2$,  (\ref{eq:LemmaMSE}) holds by plugging the required components in (\ref{eq:app2}), and implementing simple algebraic manipulations.

\section{Proof of Theorem \ref{Th1}}
\label{sec:App_Th_1}

Define $\Tilde{\eta}  \triangleq K  \eta_l$, and let $\hat{\myVec{\theta}}^{r}_{\text{MMSE}} = \myVec{\theta}^{r} + \myVec{e}_{\theta}^r$ with $\mathbb{E}[||\myVec{e}_{\theta}^r||^2] \leq \frac{d}{N^2 \alpha^r}\sigma_{\theta}^r$. We also define $\myVec{g}(\myVec{\theta}) \triangleq \frac{1}{N}\sum_{i=1}^N \myVec{g}_i(\myVec{\theta})$. For the ease of presentation, we denote $\hat{\myVec{\theta}}^{r} = \hat{\myVec{\theta}}^{r}_{\text{MMSE}}$, and we adopt the convention that summations are always over $k \in [K]$ or $i \in [N]$.

We start by decomposing the error into three terms: \vspace{0.2cm} \\
$\mathbb{E}[||\hat{\myVec{\theta}}^{r} - \myVec{\theta}^*||^2] 
 = \mathbb{E}[||\hat{\myVec{\theta}}^{r} - \hat{\myVec{\theta}}^{r-1} + \hat{\myVec{\theta}}^{r-1} - \myVec{\theta}^*||^2]$
\begin{align}
    \label{eq:app3}
    \nonumber = \mathbb{E}[||\hat{\myVec{\theta}}^{r-1} - \myVec{\theta}^*||^2] &+2\mathbb{E}[\langle \hat{\myVec{\theta}}^{r} - \hat{\myVec{\theta}}^{r-1},\hat{\myVec{\theta}}^{r-1} - \myVec{\theta}^*\rangle] \\
    &+ \mathbb{E}[||\hat{\myVec{\theta}}^{r} - \hat{\myVec{\theta}}^{r-1}||^2].
\end{align}
We first bound the third term on the RHS of (\ref{eq:app3}). Note that: \vspace{0.2cm} \\
$\hat{\myVec{\theta}}^{r} - \hat{\myVec{\theta}}^{r-1} =  \myVec{\theta}^{r} + \myVec{e}_{\theta}^r - \hat{\myVec{\theta}}^{r-1} 
 = \frac{1}{N} \sum \limits_i \myVec{\theta}_i^{r} - \hat{\myVec{\theta}}^{r-1} + \myVec{e}_{\theta}^r \vspace{0.2cm} \\
= \frac{1}{N} \sum \limits_i (\hat{\myVec{\theta}}^{r-1} - \eta_l \sum \limits_k \myVec{g}_i(\myVec{\theta}_{i,k-1}^r)) - \hat{\myVec{\theta}}^{r-1} + \myVec{e}_{\theta}^r \vspace{0.2cm} \\
= -\frac{\Tilde{\eta}}{NK} \sum \limits_{i,k} \myVec{g}_i(\myVec{\theta}_{i,k-1}^r)) + \myVec{e}_{\theta}^r. \vspace{0.2cm} \\
$
This implies that: \vspace{0.2cm} \\
$\mathbb{E} [||\hat{\myVec{\theta}}^{r} - \hat{\myVec{\theta}}^{r-1}||^2] = 
\mathbb{E} [||-\frac{\Tilde{\eta}}{NK} \sum \limits_{i,k} \myVec{g}_i(\myVec{\theta}_{i,k-1}^r)) + \myVec{e}_{\theta}^r||^2] \vspace{0.2cm} \\
= \mathbb{E} [||-\frac{\Tilde{\eta}}{NK} \sum \limits_{i,k} \myVec{g}_i(\myVec{\theta}_{i,k-1}^r))||^2] + \mathbb{E} [||\myVec{e}_{\theta}^r||^2] \vspace{0.2cm} \\
\leq \frac{\Tilde{\eta}^2}{N^2 K^2} \mathbb{E}[||\sum \limits_{i,k} \nabla f_i(\myVec{\theta}_{i,k-1}^r)||^2] +\frac{d}{N^2 \alpha^r}\sigma_{\theta}^r + \frac{\Tilde{\eta}^2}{NK} \sigma^2 \vspace{0.2cm} \\ 
\leq  2 \beta \Tilde{\eta}^2 (\mathbb{E}[F(\hat{\myVec{\theta}}^{r-1})]-F(\myVec{\theta}^*)) + 2\beta^2 \Tilde{\eta}^2  \vspace{0.2cm} \\  +
    \frac{1}{KN} \sum \limits_{i,k} 
    \mathbb{E}[||\myVec{\theta}_{i,k-1}^r - \hat{\myVec{\theta}}^{r-1}||^2] + \frac{d}{N^2 \alpha^r}\sigma_{\theta}^r + \frac{\Tilde{\eta}^2}{NK} \sigma^2$
\begin{equation}
    \label{eq:app4}
     \leq  2 \beta \Tilde{\eta}^2 (\mathbb{E}[F(\hat{\myVec{\theta}}^{r-1})]-F(\myVec{\theta}^*)) + 2\beta \Tilde{\eta}^2
    \delta^r + \frac{d}{N^2 \alpha^r}\sigma_{\theta}^r + \frac{\Tilde{\eta}^2 \sigma^2}{NK},
\end{equation}
where the first equality follows from the unbiasedness properties of the MMSE,  the second inequality follows by separating the mean and variance (Lemma \ref{lemma:separating}), the third inequality follows using similar steps as in
(\cite{karimireddy2020scaffold} Lemma 11), and $\delta^r \triangleq \frac{1}{KN} \sum \limits_{i,k} \mathbb{E}[||\myVec{\theta}_{i,k-1}^r- \hat{\myVec{\theta}}^{r-1}||^2]$.

We next bound the second term on the RHS of (\ref{eq:app3}): \vspace{0.2cm} \\
$2\mathbb{E}[\langle \hat{\myVec{\theta}}^{r} - \hat{\myVec{\theta}}^{r-1},\hat{\myVec{\theta}}^{r-1} - \myVec{\theta}^*\rangle] \vspace{0.2cm} \\ = 2\mathbb{E}[\langle -\frac{\Tilde{\eta}}{KN} \sum \limits_{i,k} \myVec{g}_i(\myVec{\theta}_{i,k-1}^r)) + \myVec{e}_{\theta}^r,\hat{\myVec{\theta}}^{r-1} - \myVec{\theta}^*\rangle] \vspace{0.2cm} \\
\leq \frac{2\Tilde{\eta}}{KN} \mathbb{E}[\sum \limits_{i,k} \langle \nabla f_i (\myVec{\theta}_{i,k-1}^r),\hat{\myVec{\theta}}^{r-1} - \myVec{\theta}^*  \rangle]$
\begin{equation}
\label{eq:app5}
 \leq  -2 \Tilde{\eta} (\mathbb{E}[F(\hat{\myVec{\theta}}^{r-1})] - F(\myVec{\theta}^*) + \frac{\mu}{4} \mathbb{E}[||\hat{\myVec{\theta}}^{r-1}-\myVec{\theta}^*||^2]) + 2\beta \Tilde{\eta} \delta^r,
\end{equation}
where the first inequality follows since the MMSE estimator is unbiased, and the last inequality follows using similar steps as in (\cite{karimireddy2020scaffold} Lemma 11). Separating again the mean and variance and using Lemma 8 in \cite{karimireddy2020scaffold}, we have:
\begin{equation}
    \label{eq:app6}
    3 \beta \Tilde{\eta} \delta^r \leq \frac{2\Tilde{\eta}}{3}(\mathbb{E}[F(\hat{\myVec{\theta}}^{r-1})] -F(\myVec{\theta}^*)) + \frac{\Tilde{\eta}^2 \sigma^2 }{2 K} + 18 \beta \Tilde{\eta}^3 G^2. 
\end{equation}
Plugging (\ref{eq:app4}),(\ref{eq:app5}),(\ref{eq:app6}) together with the assumption on the local step size leads to:
\begin{align}
    \label{eq:app7}
    \nonumber &\mathbb{E}[||\hat{\myVec{\theta}^r} - \myVec{\theta}^*||^2] \leq (1-\frac{\mu \Tilde{\eta}}{2})\mathbb{E}[||\myVec{\theta}^{r-1} - \myVec{\theta}^*||^2] \\ \nonumber &- \frac{\Tilde{\eta}}{3}(\mathbb{E}[F(\hat{\myVec{\theta}}^{r-1})] - F(\myVec{\theta}^*)) \\ 
    & + \Tilde{\eta}^2[\frac{\sigma^2}{KN}(1+N) + 18\beta \Tilde{\eta} G^2] + \frac{d}{N^2 \alpha^r} \sigma_{\theta}^r.
\end{align}
To bound $\alpha^r$, we first bound: \vspace{0.2cm} \\
$ \mathbb{E}[||\myVec{\theta}_i^r - \myVec{\theta}^{r-1}||^2] =
\mathbb{E}[||\hat{\myVec{\theta}}^{r-1} - \frac{\Tilde{\eta}}{K} \sum \limits_k g_i(\myVec{\theta}_{i,k-1}^r) - \hat{\myVec{\theta}}^{r-1}||^2] \vspace{0.2cm} \\ 
\leq \frac{\Tilde{\eta}^2}{K^2} \mathbb{E}[||\sum \limits_k g_i(\myVec{\theta}_{i,k-1}^r)||^2] \leq \frac{\Tilde{\eta}^2}{K^2} \cdot K \sum \limits_k \mathbb{E}[||g_i(\myVec{\theta}_{i,k-1}^r)||^2] \vspace{0.2cm} \\ \leq \Tilde{\eta}^2 \cdot M^2$, \vspace{0.2cm} \\
where we used the inequality $||\sum_{t=1}^{\tau} \myVec{r}_t ||^2 \leq \tau \sum_{t=1}^{\tau} ||\myVec{r}_t ||^2$, which holds for any multivariate sequence $\{\myVec{r}_t \}$. Thus we can bound 
\begin{equation}
\label{app:alpha_bound}
\frac{1}{\alpha_r} \leq \frac{\tilde{\eta}^2 M^2}{P}.  
\end{equation}

Plugging (\ref{app:alpha_bound}) into (\ref{eq:app7}) we get:
\begin{align}
    \label{eq:app8}
    \nonumber &\mathbb{E}[||\hat{\myVec{\theta}}^r - \myVec{\theta}^*||^2] \leq (1-\frac{\mu \Tilde{\eta}}{2})\mathbb{E}[||\hat{\myVec{\theta}}^{r-1} - \myVec{\theta}^*||^2] \\ \nonumber &- \frac{\Tilde{\eta}}{3}(\mathbb{E}[F(\hat{\myVec{\theta}}^{r-1})] - F(\myVec{\theta}^*)) \\ 
    & + \Tilde{\eta}^2[\frac{\sigma^2}{KN}(1+N) + \frac{d M^2 \sigma_{\theta}^r}{N^2 P} + 18\beta \Tilde{\eta} G^2].
\end{align}
Moving the $(\mathbb{E}[F(\hat{\myVec{\theta}}^{r-1})] - F(\myVec{\theta}^*))$ term and dividing throughout by $\frac{\Tilde{\eta}}{3}$, we get the following bound for any $\Tilde{\eta} \leq \frac{1}{8 \beta}$
\begin{align}
    \label{eq:app9}
    \nonumber & \mathbb{E}[F(\hat{\myVec{\theta}}^{r-1})] - F(\myVec{\theta}^*)  \\ \nonumber
    &\leq \frac{3}{\Tilde{\eta}} (1-\frac{\mu \Tilde{\eta}}{2})\mathbb{E}[||\hat{\myVec{\theta}}^{r-1} - \myVec{\theta}^*||^2] - \frac{3}{\Tilde{\eta}} \mathbb{E}[||\hat{\myVec{\theta}}^{r} - \myVec{\theta}^*||^2] \\
    & + 3\Tilde{\eta} [\frac{\sigma^2}{KN}(1+N) + \frac{d M^2 \sigma_{\theta}^r}{N^2 P} + 18 \beta \Tilde{\eta} G^2].
\end{align}

The final rate follows using the convexity of $f$ and unrolling the recursive bound in (\ref{eq:app9}) using Lemma \ref{lemma:convergence rate}, with weights $\omega_r = (1-\frac{\mu \Tilde{\eta}}{2})^{1-r}$.
\vspace{-0.2cm}

\vspace{-0.3cm}
\section{Proof of Theorem \ref{Th2}}
\label{sec:App_Th_2}
\vspace{-0.2cm}
Define $\Tilde{\eta}  \triangleq K  \eta_l$, and let $\hat{\myVec{\theta}}^{r}_{\text{MMSE}} = \myVec{\theta}^{r} + \myVec{e}_{\theta}^r$, $\hat{\myVec{c}}^{r}_{\text{MMSE}} = \myVec{c}^{r} + \myVec{e}_{c}^r$ with $\mathbb{E}[||\myVec{e}_{\theta}^r||^2] \leq \frac{d}{N^2 \alpha^r}\sigma_{\theta}^r$, $\mathbb{E}[||\myVec{e}_{c}^r||^2] \leq \frac{d}{N^2 \beta^r}\sigma_{c}^r$. We also define $\myVec{g}(\myVec{\theta}) \triangleq \sum_{i=1}^N \myVec{g}_i(\myVec{\theta})$. For ease of presentation, we denote $\hat{\myVec{\theta}}^{r} = \hat{\myVec{\theta}}^{r}_{\text{MMSE}}$, $\hat{\myVec{c}}^{r} = \hat{\myVec{c}}^{r}_{\text{MMSE}}$, and we adopt the convention that summations are always over $k \in [K]$ or $i \in [N]$.
Similarly as in the proof of Th.~\ref{Th1}, we compose the error into three terms: 
\begin{align}
\label{eq:app10}
\nonumber
\mathbb{E}[&||\hat{\myVec{\theta}}^{r} - \myVec{\theta}^*||^2] = \mathbb{E}[||\hat{\myVec{\theta}}^{r} - \hat{\myVec{\theta}}^{r-1} + \hat{\myVec{\theta}}^{r-1} - \myVec{\theta}^*||^2] \\
    \nonumber &= \mathbb{E}[||\hat{\myVec{\theta}}^{r-1} - \myVec{\theta}^*||^2] +2\mathbb{E}[\langle \hat{\myVec{\theta}}^{r} - \hat{\myVec{\theta}}^{r-1},\hat{\myVec{\theta}}^{r-1} - \myVec{\theta}^*\rangle] \\&+ \mathbb{E}[||\hat{\myVec{\theta}}^{r} - \hat{\myVec{\theta}}^{r-1}||^2].
\end{align}
Note that: \vspace{0.2cm} \\
$\hat{\myVec{\theta}}^{r} = \myVec{e}_{\theta}^r + \myVec{\theta}^{r}  =
\myVec{e}_{\theta}^r + \frac{1}{N} \sum_i \myVec{\theta}_i^{r} \vspace{0.2cm} \\ =
\myVec{e}_{\theta}^r + \frac{1}{N} \sum_i (\myVec{\theta}_{i,0}^{r} - \eta_l \sum_k (\myVec{g}_i(\myVec{\theta}_{i,k-1}^{r}) + \hat{\myVec{c}}^{r} - \myVec{c}_i^{r})) \vspace{0.2cm} \\ =
\myVec{e}_{\theta}^r + \frac{1}{N} \sum_i (\hat{\myVec{\theta}}^{r-1} - \eta_l \sum_k (\myVec{g}_i(\myVec{\theta}_{i,k-1}^{r}) + \myVec{c}^{r} +
\myVec{e}_{c}^{r} - \myVec{c}_i^{r-1})). \vspace{0.2cm} \\ $
This implies that: \vspace{0.2cm} \\
$
\hat{\myVec{\theta}}^{r} - \hat{\myVec{\theta}}^{r-1} \vspace{0.2cm} \\= 
\myVec{e}_{\theta}^r - \frac{\tilde{\eta}}{KN} \sum_{i,k}  (\myVec{g}_i(\myVec{\theta}_{i,k-1}^{r}) + \myVec{c}^{r} +
\myVec{e}_{c}^{r} - \myVec{c}_i^{r}) \vspace{0.2cm} \\=
- \frac{\tilde{\eta}}{KN} \sum_{i,k}  \myVec{g}_i(\myVec{\theta}_{i,k-1}^{r}) + 
 \myVec{e}_{\theta}^r - \tilde{\eta} \myVec{e}_{c}^{r}. 
$ \vspace{0.2cm} \\
Using the above, (\ref{eq:app10}) is bounded by: \vspace{0.2cm} \\
$
\mathbb{E}[||\hat{\myVec{\theta}}^{r-1} - \myVec{\theta}^*||^2] \vspace{0.2cm} \\ - \frac{2 \tilde{\eta}}{KN}\mathbb{E}[\langle \myVec{g}_i(\myVec{\theta}_{i,k-1}^{r}) + 
 \frac{\myVec{e}_{\theta}^r}{\tilde{\eta}}  - \tilde{\eta} \myVec{e}_{c}^{r}, \hat{\myVec{\theta}}^{r-1} - \myVec{\theta}^*\rangle] \vspace{0.2cm} \\ +  \tilde{\eta}^2 \mathbb{E}[||\frac{1}{KN} \sum_{i,k}\myVec{g}_i(\myVec{\theta}_{i,k-1}^{r}) + 
 \frac{\myVec{e}_{\theta}^r}{\tilde{\eta}} - \tilde{\eta} \myVec{e}_{c}^{r}||^2] $ 
\begin{align}
\label{eq:app11}
 \nonumber \leq  &\mathbb{E}[||\hat{\myVec{\theta}}^{r-1} - \myVec{\theta}^*||^2] - \frac{2 \tilde{\eta}}{KN}\mathbb{E}[\langle \nabla f_i(\myVec{\theta}_{i,k-1}^{r}) 
 ,\hat{\myVec{\theta}}^{r-1} - \myVec{\theta}^*\rangle]\\ \nonumber
 & + 
 \tilde{\eta}^2 \mathbb{E}[||\frac{1}{KN} \sum_{i,k} \nabla f_i(\myVec{\theta}_{i,k-1}^{r})||^2] + \frac{\tilde{\eta}^2 \sigma^2}{KN} \\
 &+ \frac{d}{N^2 \alpha^r}\sigma_{\myVec{\theta}}^r  + \frac{\tilde{\eta}^2 d}{N^2 \beta^r}\sigma_{\myVec{c}}^r,
\end{align} 
where the last inequality follows by separating
the mean and variance (Lemma \ref{lemma:separating}) and using the unbiasedness properties of the MMSE estimator. 
The first term on the RHS of (\ref{eq:app11}) is bounded similarly as in Eq.~(\ref{eq:app5}):
\begin{align}
\label{eq:app12}
\nonumber &\frac{2\Tilde{\eta}}{KN} \mathbb{E}[\sum \limits_{i,k} \langle \nabla f_i (\myVec{\theta}_{i,k-1}^r),\hat{\myVec{\theta}}^{r-1} - \myVec{\theta}^*  \rangle] \vspace{0.2cm} \\
 &\leq  -2 \Tilde{\eta} (\mathbb{E}[F(\hat{\myVec{\theta}}^{r-1})] - F(\myVec{\theta}^*) + \frac{\mu}{4} \mathbb{E}[||\hat{\myVec{\theta}}^{r-1}-\myVec{\theta}^*||^2]) + 2\beta \Tilde{\eta} \delta^r.
\end{align}
The second term on the RHS of (\ref{eq:app10}) is bounded similarly as in Eq.~(\ref{eq:app4}):
\begin{align}
    \label{eq:app13}
   \nonumber &\tilde{\eta}^2 \mathbb{E}[||\frac{1}{KN} \sum_{i,k} \nabla f_i(\myVec{\theta}_{i,k-1}^{r})||^2] \\  & \leq  2 \beta \Tilde{\eta}^2 (\mathbb{E}[F(\hat{\myVec{\theta}}^{r-1})]-F(\myVec{\theta}^*)) + 2\beta \Tilde{\eta}^2
    \delta^r. 
\end{align}
To bound the drift $\delta^r$ of the COBAAF algorithm, we first note that: \vspace{0.2cm} \\ 
$
\frac{1}{N} \sum_i \mathbb{E}[||\myVec{\theta}_{i,k}^r - \hat{\myVec{\theta}}^{r-1} ||^2] \vspace{0.2cm} \\
= \frac{1}{N} \sum_i \mathbb{E}[|| \myVec{\theta}_{i,k-1}^r - \eta_l \myVec{g}_i(\myVec{\theta}_{i,k-1}^r) + \eta_l \myVec{c}^r - \eta_l \myVec{e}_c^r - \eta_l \myVec{c}_i^r -  \hat{\myVec{\theta}}^{r-1} ||^2] \vspace{0.2cm} \\
\leq \frac{1}{N} \sum_i \mathbb{E}[|| \myVec{\theta}_{i,k-1}^r - \eta_l \myVec{g}_i(\myVec{\theta}_{i,k-1}^r) + \eta_l \myVec{c}^r - \eta_l \myVec{c}_i^r -  \hat{\myVec{\theta}}^{r-1} ||^2] + \frac{\eta_l^2 d}{N^2 \beta^r} \sigma_c^r. \vspace{0.2cm} \\$
By using the above equation, and following similar steps as in (\cite{karimireddy2020scaffold} Lemma 13), we have:
\begin{equation}
\label{eq:app14}    
3\beta \tilde{\eta} \delta^r \leq \frac{\tilde{\eta}}{2} \mathbb{E}[F(\hat{\myVec{\theta}}^{r-1}) - F(\myVec{\theta}^*)] + \frac{5 \tilde{\eta}^2 \sigma^2}{K} + \frac{2 \tilde{\eta}^2 d}{K N^2 \beta^r} \sigma_c^r. 
\end{equation}
Combining Eq.~(\ref{eq:app11})-(\ref{eq:app14}) we obtain a recursive equation for the COBAAF progress in one round: \vspace{0.2cm} \\
$
\mathbb{E}[\hat{\myVec{\theta}}^r - \myVec{\theta}^*] \leq (1-\frac{\mu \tilde{\eta}}{2})\mathbb{E}[\hat{\myVec{\theta}}^{r-1} - \myVec{\theta}^*] - \tilde{\eta} (\mathbb{E}[F(\hat{\myVec{\theta}}^{r-1})]-F(\myVec{\theta}^*)) \vspace{0.2cm}\\ 
 + \frac{\tilde{\eta}^2 \sigma^2}{KN} + \frac{d}{N^2 \alpha^r} \sigma_{\myVec{\theta}}^r + \frac{\tilde{\eta}^2 d}{N^2 \beta^r} \sigma_{\myVec{c}}^r + 3 \beta \tilde{\eta} \delta^r
$
\begin{align}
\label{eq:app15}
\nonumber & \leq (1-\frac{\mu \Tilde{\eta}}{2})\mathbb{E}[||\myVec{\theta}^{r-1} - \myVec{\theta}^*||^2] \\ \nonumber &- \frac{\Tilde{\eta}}{2}(\mathbb{E}[F(\hat{\myVec{\theta}}^{r-1})] - F(\myVec{\theta}^*)) \\ 
    & + \Tilde{\eta}^2\left[\frac{\sigma^2}{KN}(1+2N) + \frac{d \sigma_{c}^r}{N^2K \beta^r}(\frac{1}{2} + K)\right] + \frac{d}{N^2 \alpha^r} \sigma_{\theta}^r.
\end{align}
We bound $\alpha^r$ similarly as in (\ref{app:alpha_bound}), and $\beta^r$ can be bounded directly by $\frac{1}{\beta^r} \leq \frac{M^2}{P}$.
Arranging the terms above, we get the following bound for any $\eta_l \leq \frac{1}{2 \beta K}$:\\
$\mathbb{E}[F(\hat{\myVec{\theta}}^{r-1})] - F(\myVec{\theta}^*)$  \\
\begin{align}
    \label{eq:app16}
    \nonumber 
    &\leq \frac{2}{\Tilde{\eta}} (1-\frac{\mu \Tilde{\eta}}{2})\mathbb{E}[||\hat{\myVec{\theta}}^{r-1} - \myVec{\theta}^*||^2] - \frac{2}{\Tilde{\eta}} \mathbb{E}[||\hat{\myVec{\theta}}^{r} - \myVec{\theta}^*||^2] \\
    & + 2\Tilde{\eta} \left[\frac{\sigma^2}{KN}(1+2N + \frac{dM^2 \sigma_{\theta}}{NP\sigma^2}) + \frac{dM^2 \sigma_{c}^r}{N^2K P}(K + 2) \right]. 
\end{align}
The final rate follows using the convexity of $f$ and unrolling the recursive bound in (\ref{eq:app16}) using Lemma \ref{lemma:convergence rate}, with weights $\omega_r = (1-\frac{\mu \Tilde{\eta}}{2})^{1-r}$.
\vspace{-0.3cm}
\section{Proof of Theorem \ref{Th3}}
\label{sec:App_Th_3}
\vspace{-0.2cm}
With the same definitions as in Appendix \ref{sec:App_Th_2} (and this time we adopt the convention that the summations of the user index are always over $i \in \mathcal{S}^r$) we compose the error into three terms:
\begin{align}
\label{eq:app17}
\nonumber
\mathbb{E}[&||\hat{\myVec{\theta}}^{r} - \myVec{\theta}^*||^2] = \mathbb{E}[||\hat{\myVec{\theta}}^{r} - \hat{\myVec{\theta}}^{r-1} + \hat{\myVec{\theta}}^{r-1} - \myVec{\theta}^*||^2] \\
    \nonumber &= \mathbb{E}[||\hat{\myVec{\theta}}^{r-1} - \myVec{\theta}^*||^2] +2\mathbb{E}[\langle \hat{\myVec{\theta}}^{r} - \hat{\myVec{\theta}}^{r-1},\hat{\myVec{\theta}}^{r-1} - \myVec{\theta}^*\rangle] \\&+ \mathbb{E}[||\hat{\myVec{\theta}}^{r} - \hat{\myVec{\theta}}^{r-1}||^2].
\end{align}
Here we have
\begin{equation}
\label{eq:app18}   
\hat{\myVec{\theta}}^{r} - \hat{\myVec{\theta}}^{r-1} = -\frac{\tilde{\eta}}{KS} \sum \limits_{i,k} (g_i(\myVec{\theta}_{i,k-1}) + \hat{\myVec{c}}^r - \myVec{c}_i) + e_{\theta}^r
\end{equation}
and 
$\mathbb{E}[\hat{\myVec{\theta}}^{r} - \hat{\myVec{\theta}}^{r-1}] =-\frac{\tilde{\eta}}{KN} \sum \limits_{i,k} g_i(\myVec{\theta}_{i,k-1}).$
Separating mean and variance, we can bound:
\begin{align}
\label{eq:app20} \nonumber  
\mathbb{E}[||\hat{\myVec{\theta}}^{r} - \hat{\myVec{\theta}}^{r-1}||^2] &\leq \mathbb{E}[||\frac{\tilde{\eta}}{KN} \sum \limits_{i,k} (g_i(\myVec{\theta}_{i,k-1}) + \myVec{c}^r - \myVec{c}_i)||^2]  \\+&\frac{d}{N^2 \alpha^r}\sigma_{\myVec{\theta}}^r  + \frac{\tilde{\eta}^2 d}{N^2 \beta^r}\sigma_{\myVec{c}}^r.
\end{align}
Using Lemma $11$ from \cite{karimireddy2020scaffold}, we have:
\begin{align}
\label{eq:app21}
    \nonumber &\mathbb{E}[||\frac{\tilde{\eta}}{KN} \sum \limits_{i,k} (g_i(\myVec{\theta}_{i,k-1}) + \myVec{c}^r - \myVec{c}_i)||^2] 
    \\ &\leq 8\beta \tilde{\eta}^2 (\mathbb{E}[f(\hat{\myVec{\theta}})]-f(\myVec{\theta^*})) + 8\tilde{\eta}^2 \mathcal{C}^{r} + 4\tilde{\eta}^2 \beta^2 \delta^r + \frac{12\tilde{\eta}^2\sigma^2}{KS},
\end{align}
where
\begin{equation}
\label{eq:app22}
\mathcal{C}^{r} \triangleq \frac{1}{N} \sum \limits_{i=1}^N \mathbb{E}[||\mathbb{E}[\myVec{c}_i^r]-\nabla f_i(\myVec{\theta}^*)||^2]
\end{equation}
is the "control-lag" as discussed in Section \ref{ssec:fading_channels}.
We now bound $\mathcal{C}^{r}$ recursively using Lemma $12$ in \cite{karimireddy2020scaffold}:
\begin{equation}
\label{eq:app23}
\mathcal{C}^{r} \leq (1-\frac{S}{N})\mathcal{C}^{r-1} + \frac{S}{N}(4\beta(\mathbb{E}[f(\hat{\myVec{\theta}}^{r-1})] - f(\myVec{\theta}^*)) + 2\beta^2 \delta^r).
\end{equation}
The drift can be bound similarly as in Appendix \ref{sec:App_Th_2}, with the addition of the control-lag parameter:
\begin{align}
\label{eq:app24}   
\nonumber
3\beta \tilde{\eta} \delta^r &\leq \frac{\tilde{\eta}}{2} \mathbb{E}[F(\hat{\myVec{\theta}}^{r-1}) - F(\myVec{\theta}^*)] + \frac{5 \tilde{\eta}^2 \sigma^2}{K} + \frac{2 \tilde{\eta}^2 d}{K N^2 \beta^r} \sigma_c^r \\&+ \frac{2\tilde{\eta}^2}{3}\mathcal{C}^{r-1}. 
\end{align}
The same arguments used in proving Theorem \ref{Th2} can now be applied to prove the theorem.

\bibliographystyle{ieeetr}
\bibliography{Ga_Co_EL_Bayesian}

\end{document}

%% file: macros.tex
\newtheorem{theorem}{Theorem}
\newtheorem{lemma}{Lemma}

\newtheorem{definition}{Definition}

\newcommand{\beq}{\begin{equation}}
\newcommand{\eeq}{\end{equation}}
\newcommand{\bea}{\begin{array}}
\newcommand{\ena}{\end{array}}
\newcommand{\bds}{\begin {itemize}}
\newcommand{\eds}{\end {itemize}}
\newcommand{\bdf}{\begin{definition}}
\newcommand{\blm}{\begin{lemma}}
\newcommand{\edf}{\end{definition}}
\newcommand{\elm}{\end{lemma}}
\newcommand{\bthm}{\begin{theorem}}
\newcommand{\ethm}{\end{theorem}}
\newcommand{\bprp}{\begin{prop}}
\newcommand{\eprp}{\end{prop}}
\newcommand{\bcl}{\begin{claim}}
\newcommand{\ecl}{\end{claim}}
\newcommand{\bcr}{\begin{coro}}
\newcommand{\ecr}{\end{coro}}
\newcommand{\bquest}{\begin{question}}
\newcommand{\equest}{\end{question}}

\newcommand{\larrow}{{\larrow}}

\def\urltilda{\kern -.15em\lower .7ex\hbox{\~{}}\kern .04em}